\documentclass[%
 reprint,
 nofootinbib,
 nobibnotes,
 amsmath,amssymb,
 aps,
prl]{revtex4-2}

\usepackage{graphicx}% Include figure files
\usepackage{dcolumn}% Align table columns on decimal point
\usepackage{bm}% bold math
\usepackage{siunitx}
\usepackage{amsmath,amssymb}
\usepackage{graphicx}
\usepackage{textcomp,gensymb}
\usepackage[symbol,stable]{footmisc}
\usepackage{booktabs}
\usepackage{makecell}
\usepackage{threeparttable}
\usepackage{array}
\newcolumntype{P}[1]{>{\centering\arraybackslash}p{#1}}
\usepackage{amsthm}
\usepackage{mathtools}
\usepackage{csquotes}
\usepackage{tikz}
\usepackage{multirow}
\usepackage{footnote}

\begin{document}

\preprint{APS/123-QED}

\title{Volumetric Nanophotonics for Enhanced Scintillation and Remote Light Outcoupling}

\author{Simo~Pajovic$^{\bigstar,1,\ddagger}$}
\email{pajovics@mit.edu}
\author{Marius~Jürgensen$^{\bigstar,2,3}$}
\author{Sachin~Vaidya$^{4, 5}$}
% \email{svaidya1@mit.edu}
\author{Johnathon~P.~Gales$^{3}$}
\author{Joshua~Chen$^{5}$}
\author{Steven~E.~Kooi$^{6}$}
\author{Sion~Richards$^{7}$}
\author{Isabel~H.~B.~Braddock$^{7}$} 
\author{Karen~Mamian$^{8}$}
\author{William~Michaels$^{5}$}
\author{Shaul~Katznelson$^{8,9}$}
\author{Chris~D.~Armstrong$^{7}$}
\author{Ido~Kaminer$^{9}$}
\author{Marin~Soljačić$^{4, 5}$}
\author{Mikael~C.~Rechtsman$^{3}$}
\author{Charles~Roques-Carmes$^{5,8,10}$}
\email{crc@ista.ac.at}

\affiliation{\vspace{7pt}$^{1}$Department of Mechanical Engineering, Massachusetts Institute of Technology, Cambridge, MA, USA\looseness=0}
\affiliation{$^{2}$Department of Physics, Stanford University, Stanford, CA, USA\looseness=-1}
\affiliation{$^{3}$Department of Physics, The Pennsylvania State University, University Park, PA, USA\looseness=-1}
\affiliation{$^{4}$Department of Physics, Massachusetts Institute of Technology, Cambridge, MA, USA\looseness=-1}
\affiliation{$^{5}$Research Laboratory of Electronics, Massachusetts Institute of Technology, Cambridge MA USA\looseness=-1}
\affiliation{$^{6}$Institute for Soldier Nanotechnologies, Massachusetts Institute of Technology, Cambridge, MA, USA\looseness=0}
\affiliation{$^{7}$Science and Technology Facilities Council, Rutherford Appleton Laboratory, Didcot, UK\looseness=-1}
\affiliation{$^{8}$Institute of Science and Technology Austria (ISTA), Klosterneuburg, Austria\looseness=-1}
\affiliation{$^{9}$Solid State Institute, Technion - Israel Institute of Technology, Haifa, Israel\looseness=-1}
\affiliation{$^{10}$E. L. Ginzton Laboratory, Stanford University, Stanford, CA, USA\looseness=-1}
\affiliation{$^{\ddagger}$Current address: Department of Applied Physics and Materials Science, California Institute of Technology, Pasadena, CA, USA\looseness=0\\
$^{\bigstar}$ denotes equal contribution.}

               % but any date may be explicitly specified

\begin{abstract}
% The light yield of a scintillator is an intrinsic material property. However, the overall efficiency of a scintillating device, such as a detector, depends on how much of the scintillated light can be extracted. 
Nanophotonic patterning is a promising approach to enhancing light extraction in scintillator materials, with the potential to improve performance in applications such as medical imaging and particle detection. While 2D surface patterns have been successful in improving scintillator performance, the true potential of nanophotonic patterning lies in the bulk, where photons are emitted, transported, and absorbed. Here, we experimentally demonstrate the enhancement and control of scintillation using 2.5D volumetric nanophotonic patterns. Using a direct femtosecond laser writing approach, we nanofabricate square lattices of air holes into the bulk of inorganic scintillators (YAG:Ce and YAG:Yb), achieving nearly subwavelength periods compared to the peak emission wavelength and depths up to hundreds of microns, which fill a large fraction of the full thickness of the bulk. Under high-energy excitation, the patterns appear up to $\sim3.5\times$ brighter and demonstrate emission enhancement trends consistent with well-established nanophotonic scaling laws. We also report a novel functionality we call ``remote outcoupling,'' in which the nanophotonic patterns emit light that is generated in another region of the scintillator, without being directly excited themselves. Our monolithic nanofabrication approach extends nanophotonic scintillators into the bulk, which is an important step towards total control over scintillation using photonic design.
\end{abstract}

%Use showkeys class option if keyword
\maketitle

%\tableofcontents

\section*{Introduction}

Scintillators are critically important materials for the indirect detection of ionizing radiation in applications ranging from medical and industrial X-ray imaging to security screening and high-energy particle detection. In all these systems, performance is constrained by fundamental, material-level tradeoffs between stopping power, light yield, decay time, afterglow, and radiation transport, which ultimately limit sensitivity, timing, and spatial resolution~\cite{Lecoq2017,Wang2023,Singh2024}. Recently, nanophotonic scintillators have been proposed as an alternative route to bypass these tradeoffs, achieved by manipulating the dielectric environment at the length scale of the emission wavelength to control light extraction, the local density of optical states, directionality of emitted light, and spectral matching to the detector~\cite{Kurman2020,RoquesCarmes2022,Ye2022,RoquesCarmes2023,Shultzman2023,MartinMonier2024, Shultzman2024,Kurman2024,Long2024,Makowski2025}.

The vast majority of previous work on nanophotonic scintillators has focused on relatively shallow, 2D surface patterning~\cite{Knapitsch2014,Zhu2015b,Zhu2015a,Zhu2015c,Zanettini2016,Liu2017,Zhang2017,Chen2018,Salomoni2018,RoquesCarmes2022, MartinMonier2024,Zabel2026}. To realize the full potential of these systems, however, the nanophotonic control of light emission must move beyond the surface and deep into the bulk. This is challenging due to an extreme length scale mismatch: while nanophotonically-driven surface effects already occur for wavelength-deep patterns, on the order of a few hundred nanometers or less, real-world scintillators are typically a few hundreds of microns to centimeters thick~\cite{Lecoq2017}. Even the thinnest scintillators used in synchrotrons, which are a few microns thick, are already an order of magnitude larger than the depth of most surface patterns~\cite{Wollesen2022}. As a result of this length scale mismatch, most nanophotonic scintillators that have been experimentally demonstrated so far affect light emission only at the surface and completely ignore the bulk, where the majority of photonic processes such as emission, absorption, and transport take place. Overcoming this gap could open the door to complete, three-dimensional control of scintillation, including emission rate, directionality, transport, and light extraction throughout the active medium. In turn, this could ultimately enable thick scintillators with simultaneously improved efficiency, brightness, and spatial resolution~\cite{Choi2026a}.

While volumetric nanostructuring has remained a longstanding fabrication challenge in all areas of photonics, recent progress has started to bring wavelength-scale, bulk patterning to conventional optical materials~\cite{Oran2018,Rodenas2019,Han2022,Sabet2024,Zhang2024,PazBuclatin2025,PazBuclatin2026}. In scintillators, internal micro- and mesoscale architectures---including columnar CsI and laser-induced optical barriers---have already shown that volumetric structuring can mitigate the enduring tradeoff between absorption efficiency and spatial resolution, albeit at feature sizes still far larger than the subwavelength, nanophotonic regime~\cite{Lin2022,Sabet2016}. Micron-scale multilayer structures an order of magnitude thicker than the emission wavelength have achieved Purcell-enhanced scintillation, but they are challenging to scale up beyond a few microns~\cite{Kurman2024,Beer2025}. Recently, self-assembled nanoplasmonic perovskites were used to demonstrate that Purcell-enhanced scintillation can be extended into the bulk~\cite{Makowski2025}. Wavelength-scale volumetric nanophotonic patterns have not been realized directly inside an inorganic scintillator, yet closing this gap would make it possible to engineer light-matter interactions throughout the full, active volume of scintillators or other light-emitting material platforms. 
% This could then open the door for thick scintillators with enhanced emission and spatial resolution, as well as the potential for entirely new functionalities enabled by uniquely 3D nanophotonic effects.

Here, we experimentally demonstrate a platform for volumetric nanophotonic scintillators. The structures are fabricated in YAG:Ce and YAG:Yb using femtosecond laser writing and subsequent wet-etching, a technique initially developed for laser crystals~\cite{Rodenas2019}. We realize 2.5D periodic nanostructures---meaning 2D extruded into the third dimension---with nearly subwavelength-scale lateral periods and depths of hundreds of microns. Our structures show enhanced brightness and control scintillation beyond the scope of what is possible with surface patterning only. We show that scintillation can be tuned via the pattern period and etch depth, and we uncover a remote outcoupling phenomenon in which patterned regions of the scintillator capture and emit scintillated photons generated remotely in the bulk, without being directly excited themselves.

\section*{Results}

\begin{figure*}
\includegraphics{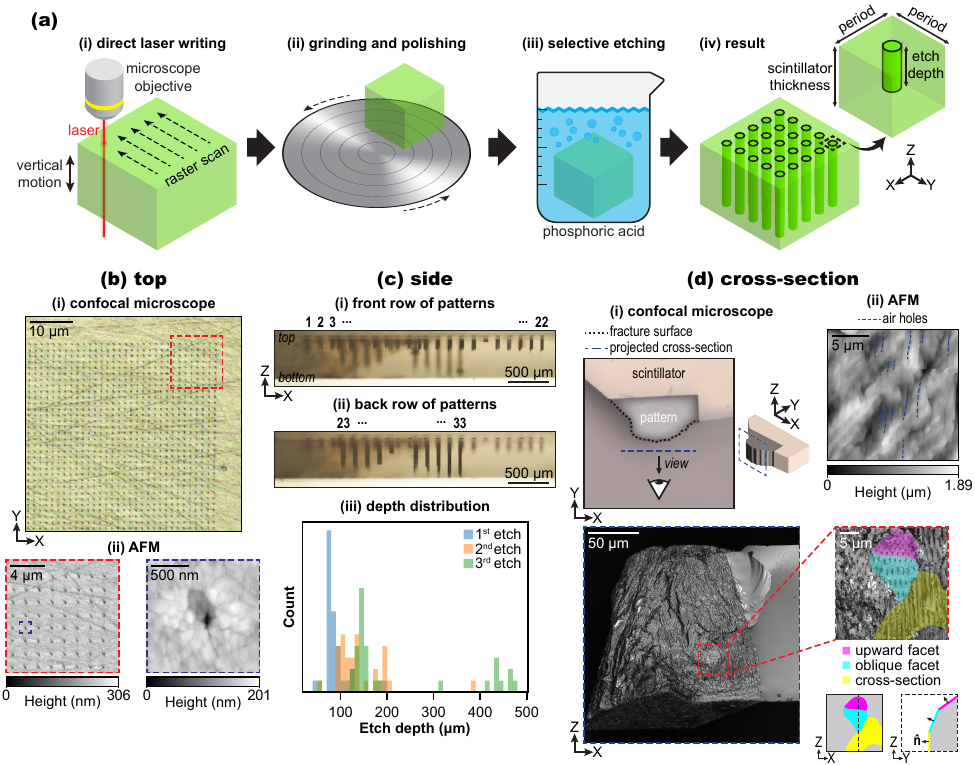}
\caption{\small \textbf{Fabrication and characterization of volumetric nanophotonic scintillators.} (a) We achieve volumetric nanophotonic patterning in three steps: (i) direct femtosecond laser writing, (ii) grinding and polishing, and (iii) selective etching in phosphoric acid. (iv) shows the resulting volumetric nanophotonic scintillator, defined by its period and etch depth, which can be comparable to its thickness. (b) A typical volumetric nanophotonic scintillator, viewed from the top using (i) confocal microscopy and (ii) atomic force microscopy (AFM). The sample shown is PC-16 (YAG:Ce, see Supplementary Materials S4 for further details). (c) Side optical microscope view of a volumetric nanophotonic scintillator. The sample shown is PC-27 (YAG:Yb). These images can be used to estimate the etch depths of the patterns. The histogram shows the etch depth distribution aggregated over 33 patterns on the same bulk scintillator as a function of the number of five-hour-long etching sessions. This shows that material is gradually removed over three etching sessions. (d) Cross-sectional view of a similar pattern (sample PC-14, YAG:Ce). The bulk scintillator shown in the confocal microscope image (i) fractured through a pattern (see Supplementary Materials S4, Fig. S10). In the enlargement of the confocal microscope image, the facets making up the fractured surface are indicated by different colors. The cross-section of the air holes, showing that they are indeed etched all the way through, is highlighted by the yellow region. This is supported by the AFM image (ii), in which air holes are highlighted by the blue dashed lines.}
\label{fig:fabrication_characterization}
\end{figure*}

We fabricate volumetric nanophotonic scintillators using direct femtosecond laser writing followed by selective etching, as illustrated in Fig.~\ref{fig:fabrication_characterization}(a). This technique was adapted from a method developed for laser crystals such as Nd:YAG~\cite{Rodenas2019}. High-power laser pulses are focused into the scintillator and chemically modify the material at the focal point of the microscope objective. By vertically moving the scintillator on a translation stage, as shown in Fig.~\ref{fig:fabrication_characterization}(a)(i), millimeter deep features can be written. The feature size is ultimately limited by the working distance of the microscope objective. Repeating this step at different transverse positions (raster scanning) allows us to write 2.5D, volumetric nanophotonic patterns. After writing, the scintillator is ground and polished to expose the written regions, which are then selectively etched using phosphoric acid, Fig.~\ref{fig:fabrication_characterization}(a)(ii)--(iii). The result is a volumetric nanophotonic pattern, consisting of air holes with giant aspect ratios that penetrate hundreds to thousands of wavelengths deep into the scintillator. Figure~\ref{fig:fabrication_characterization}(a)(iv) shows that the patterns are defined by their period and etch depth, corresponding to the chemically-modified material removed by selective etching. Note that the etch depth may not be the same as the written or targeted depth if the hole was not etched all the way through.

Our results synthesize experimental observations and data from several different samples. Each sample is a bulk scintillator with one or more volumetric patterns fabricated into it. Since bulk scintillators of the same variety (YAG:Ce or YAG:Yb) were sourced from the same suppliers and the nanofabrication method was the same for all samples, the generality of the conclusions supported by our results should be largely unaffected by individual differences between the samples. In what follows, the name of each sample will be stated in the relevant figure captions, and we refer readers to Supplementary Materials S4 for further details.

Figure~\ref{fig:fabrication_characterization}(b) shows a top view of a typical pattern nanofabricated into YAG:Ce. The pattern has a well-defined period, and the holes are clearly visible. We estimate the etch depth by imaging the patterns from the side using optical microscopy (see Supplementary Materials S3A for details). Figure \ref{fig:fabrication_characterization}(c) shows two example images with several patterns (periods ranging from 1.0--1.5~\unit{\micro\meter}) next to and behind each other after 15 h of etching. In particular, the patterns numbered 31--33 (Fig.~\ref{fig:fabrication_characterization}(c)(ii)) are etched almost all the way through the bulk scintillator. The patterns shown were thrice-etched, with imaging between each five-hour-long etching session. In Fig.~\ref{fig:fabrication_characterization}(c)(iii), we plot the estimated etch depths of all 33 patterns after each etching session. This data demonstrates that, for the most part, repeated etching removes material and makes the patterns deeper.

To complete our characterization, Fig. \ref{fig:fabrication_characterization}(d) shows the cross-section of a typical pattern, which was serendipitously obtained when the sample fractured through one of the patterns. Note that since the bulk scintillators are thin ($\sim$0.5--1~\unit{\milli\meter}) and brittle, it is difficult to intentionally mill or otherwise expose the interior structure of a pattern without destroying it. As can be seen, the fractured surface is rough, having many randomly-oriented facets. In the enlargement of the confocal microscope image in Fig.~\ref{fig:fabrication_characterization}(d)(i), we interpret the data as follows: in the yellow shaded region, the vertical cross-sections of the holes (along their depths) appear as dark, periodic lines aligned with the holes in the magenta and blue shaded regions. An AFM image of the fractured surface, shown in Fig.~\ref{fig:fabrication_characterization}(d)(ii), provides additional evidence that the holes can be observed inside the bulk scintillator, indicated by the blue dashed lines. These images show that the holes extend deep into the structure while retaining their order and morphology, serving as evidence that the patterns are truly 2.5D, volumetric nanophotonic patterns. Importantly, the results of our characterization show that these patterns are tunable in depth while remaining ordered and periodic even deep inside the bulk scintillator. Nevertheless, some amount of inescapable fabrication disorder is expected and may even enable useful scattering-driven effects, to be discussed.

\section*{Nanophotonic Scintillation Enhancement}

We expect that the volumetric nanophotonic patterns will enhance scintillation by more efficiently outcoupling light from the bulk into free space, similarly to prior work~\cite{RoquesCarmes2022,MartinMonier2024}. This material platform allows us to measure scintillation enhancement---here referring to an enhancement in the directional intensity of scintillated light collected by our microscope objective---as a function of period as well as etch depth. Scaling laws have already established the dependence of scintillation enhancement on the period~\cite{Yu2010pnas}, but a clear and potentially meaningful etch depth dependence has not been observed until now, especially as the etch depth extends well beyond the wavelength-scale regime.

\begin{figure*}
\centering
\includegraphics{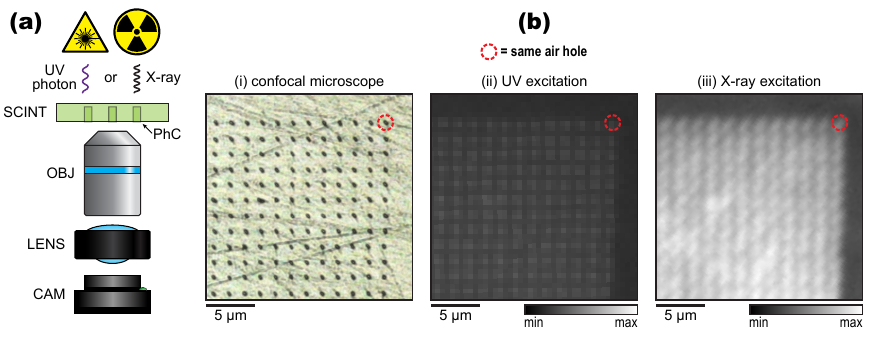}
\caption{\textbf{Measuring scintillation enhancement.} (a) Experimental setup. A scintillator is excited by UV or X-ray photons and directly imaged using a high-NA objective. (b) We observed light emission from the individual air holes that make up each volumetric nanophotonic pattern. The signal coming from the holes can be correlated across (i) confocal microscope (no excitation), (ii) UV excitation, and (iii) X-ray excitation images, indicated by the dashed red circle. The sample shown is PC-16 (see Supplementary Materials S4 for further details).}
\label{fig:air_holes}
\end{figure*}

Figure \ref{fig:air_holes}(a) schematizes our experimental setup. The sample, which has multiple patterns on it (labeled ``PhC,'' for ``Photonic Crystal''), is illuminated by either a UV laser (Spectra-Physics) or the source of a commercially available micro-CT (ZEISS), causing it to emit light. This is collected by a large numerical aperture (NA) objective (Nikon, 0.45 NA) and a tube lens (Thorlabs) to form an image on a camera (Hamamatsu, Andor). Figure \ref{fig:air_holes}(b) shows three example images of the same region of interest of the same pattern: (i) a confocal microscope image (no excitation), (ii) UV excitation, and (iii) X-ray excitation.

Interestingly, the UV and X-ray excitation images exhibit a spatial modulation that follows the periodicity of the pattern, with bright spots appearing in the vicinity of each individual air hole. The apparent correspondence between unit cells in the different images (i)--(iii) is indicated by the dashed red circles. This observation demonstrates that the nanostructure can control the spatial distribution of scintillation and is consistent with localized light extraction or scattering near the air holes. However, the finite imaging resolution of our experimental setup prevents us from determining whether the bright spots originate within the holes or their immediate surroundings, or from identifying the underlying mechanism. In particular, the observed spatial modulation may be a manifestation of disorder-induced scattering and/or an enhancement of the local density of optical states (LDOS). Distinguishing between these mechanisms would require experiments involving more localized excitation and improved spatial resolution, together with control over the pattern period, etch depth, surface roughness, and fabrication disorder.

\begin{figure*}
\centering
\includegraphics{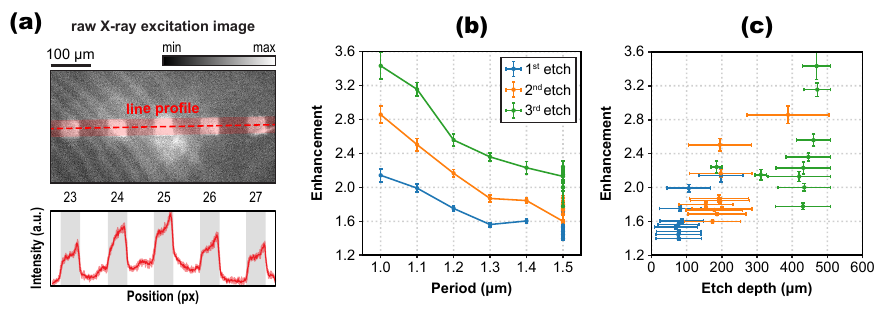}
\caption{\textbf{Volumetric nanophotonic scintillation enhancement.} (a) Measuring scintillation enhancement using line profiles drawn across the patterns. The intensity line profile shown is not exactly to scale, but it is the true line profile for the patterns in the example image (note that the image is cropped). The sample shown is PC-27, which is made of YAG:Yb (emission wavelength 1030~\unit{\nano\meter}, see Supplementary Materials S4 for further details). (b) Scintillation enhancement is estimated using Eqs. (\ref{eq:spatial_enhancement}) and (\ref{eq:average_enhancement}) and plotted as a function of period. (c) Same as (b), but plotted as a function of etch depth. The legend in (b) applies to (c) as well.}
\label{fig:emission_enhancement}
\end{figure*}

The measurement of emission enhancement under X-ray excitation is illustrated in Fig. \ref{fig:emission_enhancement}(a) using an example image captured by the camera. A line profile is drawn across the image, including all the patterns in the field of view, as shown by the red dashed line. The signal is averaged over the width of the line profile, indicated by the red shaded region. Each pattern is labeled with a known number correlated with a reference optical microscope image (see Fig. \ref{fig:fabrication_characterization}(c), for example, or Supplementary Materials S4). An example line profile is plotted below the example image as a function of pixel position and raw intensity.

The scintillator shown in Fig. \ref{fig:emission_enhancement}(a) is YAG:Yb, which emits at a peak wavelength of 1030~\unit{\nano\meter}. This wavelength can be difficult to detect because many silicon-based cameras have poor quantum efficiency ($<10$\%) in this part of the near-infrared. Even with the best camera available to us, high-energy and high-power X-ray photon excitation, and pixel binning, the signal was weak and etaloning was observed in the images (the dark, undulating bands in Fig. \ref{fig:emission_enhancement}(a))~\cite{etaloning}. To correct for etaloning and signal fluctuations, the measurements were carried out as follows: first, we obtained three line profiles to measure the intensity of light emitted by the patterns. Figure~\ref{fig:emission_enhancement}(a) shows an example of a line profile averaged from these three measurements, with the shaded area as error bars. Then, we moved the patterns away from the middle of the image and obtained three line profiles measuring the intensity of the unpatterned scintillator. After repeating this process for all 33 patterns, we measured the background intensity using the same procedure. The thrice-repeated measurements were averaged, and we calculated the spatially-dependent emission enhancement (i.e., as a function of line profile position $x$) using the following formula:
\begin{equation}
    \eta'(x)=\frac{\overline{P}(x)-\overline{B}(x)}{\overline{U}(x)-\overline{B}(x)},
\label{eq:spatial_enhancement}
\end{equation}
where $\overline{P}(x)$, $\overline{U}(x)$, and $\overline{B}(x)$ are the average, spatially-dependent patterned, unpatterned, and background intensities, respectively. By calculating the enhancement on a spatial basis, the etaloning present in all signals cancels out. Since $\eta'(x)$ jumps whenever a pattern is present, we used an edge detection algorithm to find the edges of the patterns, $x_{i,L}$ and $x_{i,R}$ (where $L$ and $R$ denote the left and right edges, respectively, and $i$ denotes the pattern number). Finally, the scintillation enhancement of each pattern $i$ can be calculated as
\begin{equation}
    \eta_{i} = \frac{1}{x_{i,R}-x_{i,L}}\int_{x_{i,L}}^{x_{i,R}}\eta'(x)dx.
\label{eq:average_enhancement}
\end{equation}

This approach, which is described in further detail in Supplementary Materials S3B, allows us to measure scintillation enhancement as a function of period and etch depth, since we can measure scintillation enhancement after each session of etching. These trends are plotted in Fig.~\ref{fig:emission_enhancement}(b) and (c) for the YAG:Yb scintillator also shown in Fig.~\ref{fig:fabrication_characterization}(c). Firstly, scintillation enhancement increases as the period shrinks and approaches the subwavelength regime (in this case, the free-space peak emission wavelength, $\lesssim1030$~\unit{\nano\meter}). This is true regardless of the etching session (indicated by the different colors), which is a proxy for etched depth vis-à-vis Fig.~\ref{fig:fabrication_characterization}(c)(iii). Scintillation enhancement approaching the subwavelength regime is a well-known prediction from density of states arguments for 2D photonic crystals~\cite{Yu2010pnas,RoquesCarmes2022}. Our results demonstrate that these trends can hold for volumetric photonic crystals just as well. In this case, the patterns can be thought of as 2.5D photonic crystals since the holes are uniform in the $z$-direction. Notably, scintillation enhancement also increases after each etch round, suggesting that increasing etch depth increases scintillation enhancement. This is further demonstrated in the adjacent plot, where scintillation enhancement is directly plotted against etch depth (ignoring the period of each pattern). In spite of the large horizontal error bars associated with the estimated etch depth (resulting from blur in the optical microscope images in Fig. \ref{fig:fabrication_characterization}(c), see Supplementary Materials S3A for details), the general trend is clear: scintillation enhancement increases with the etched depths of the patterns.

\section*{Remote Outcoupling of Scintillation}

\begin{figure*}
\includegraphics{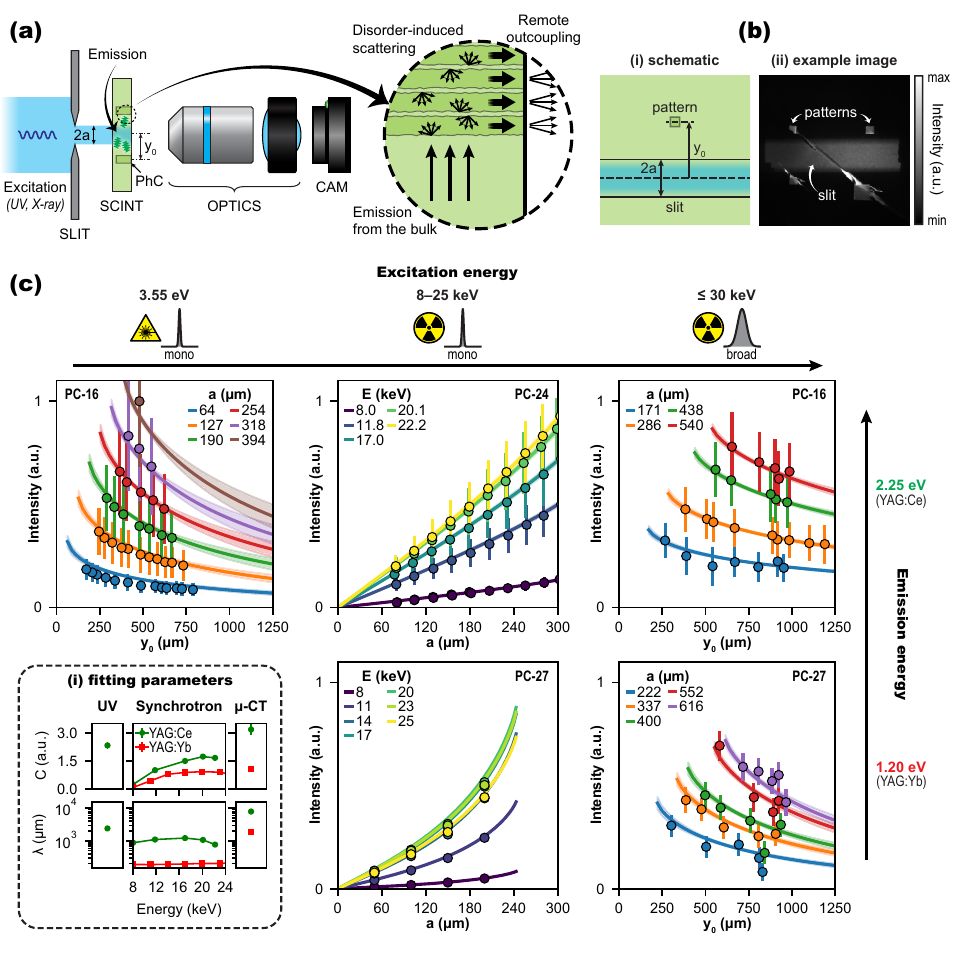}
\caption{\small \textbf{Remote outcoupling of light in volumetrically-patterned nanophotonic scintillators.} (a) Experimental setup. The bulk scintillator, imaged using a camera, is illuminated by a source of high-energy photons (UV or X-ray), which are partly blocked by a slit of width $2a$. The center of each pattern (``PhC'') with respect to the center of the slit is $y_0$, and there may be patterns on either side of the slit. In contrast to Fig.~\ref{fig:emission_enhancement}, the patterns are not directly excited by the source, so any light emission by the patterns should be mostly due to remote outcoupling of light. (Inset) Laterally-propagating light emitted in the slit region is intercepted by the patterns. This light is scattered in all directions because of fabrication disorder, resulting in measurable ``remote outcoupling'' along the holes toward the camera. (b) Our simple model's setup along with an example image of a YAG:Ce scintillator showing the slit in the center and the relatively bright patterns above and below it. The sample shown is PC-16 (Supplementary Materials S4). (c) Measured, pattern-averaged remote outcoupling (markers with error bars) for two different scintillators (YAG:Ce, $E_0\sim2.25$~\unit{\eV}, and YAG:Yb, $E_0\sim1.20$~\unit{\eV}) excited by three different sources: a 3.55~\unit{\eV} (349~\unit{\nano\meter}) UV laser, a synchrotron ranging from 8--25~\unit{\kilo\eV}, and an X-ray tube with a maximum X-ray energy of 30~\unit{\kilo\eV}. The relevant sample names are written inside the axes across from the legends for cross-referencing with Supplementary Materials S4. In the cases of the UV laser and the X-ray tube, intensity is plotted against $y_0$ for a number of different $a$-values, while in the case of the synchrotron, it is plotted against $a$ for a number of different excitation energies. Each plot represents an independent set of measurements, with the intensities normalized to their maximum value in their respective datasets. The solid lines are the fits to Eq. (\ref{eq:dipole_model}), with the shaded areas representing the 95\% confidence interval. (i) shows the fitting parameters in Eq. (\ref{eq:dipole_model}) as a function of the excitation source or energy, showing consistent general trends between scintillators.}
\label{fig:light_collecting}
\end{figure*}

In addition to scintillation enhancement, we observed a functionality of the volumetric patterns that, to the best of our knowledge, has not been previously reported. When the patterns are not directly excited, but another region of the scintillator is (e.g., by using an aperture or slit to block part of the source), the patterns still emit light. We hypothesize that the patterns intercept laterally propagating light emitted relatively far away in the bulk scintillator and, via scattering due to the presence of disorder and the resulting lack of momentum conservation at the interface, remotely outcouple the emitted light in the transverse direction. Our hypothesis, along with the experimental setup we built to investigate this remote outcoupling functionality, is shown in Fig. \ref{fig:light_collecting}(a). The setup is nearly identical to the one used to measure scintillation enhancement except that the source (again, UV or X-ray photons, though in this case the X-ray photons can be either monochromatic from a synchrotron or broadband from an X-ray tube) is partly blocked by a slit of width $2a$. This excites a finite, rectangular region of the bulk scintillator that is located a distance $y$ from the patterns. This is further illustrated in Fig. \ref{fig:light_collecting}(b), which shows an example image in which the patterns are not directly excited by UV light but appear relatively bright.

To understand this remote outcoupling effect in volumetric nanophotonic scintillators, we formulate a simple model in two dimensions. Fig.~\ref{fig:light_collecting}(b) shows a schematic of our simple model's setup compared to an example image from our experiments. We treat the slit or excitation region as a uniform distribution of isotropically-emitting dipoles infinite along one direction. Summing up the contributions from all the dipoles leads to the following equation for the intensity of light at the center $y_0$ of a given pattern:
\begin{widetext}
\begin{equation}\label{eq:dipole_model}
    I_E(a,y_0)=C(E)[r_{+}L_{-1}(r_{+})K_0(r_{+}) + r_{+}L_0(r_{+})K_1(r_{+}) - r_{-}L_{-1}(r_{-})K_0(r_{-}) - r_{-}L_0(r_{-})K_1(r_{-})],
\end{equation}
\end{widetext}
\noindent where $r_{\pm}=(y_0 \pm a)/\lambda$, $K_n$ are modified Bessel functions of the second kind, and $L_n$ are modified Struve functions~\cite{AbramowitzStegun}. The full derivation of our simple model can be found in Supplementary Materials S5. The parameters $C(E)$ and $\lambda$ can be fitted, with $\lambda$ being interpreted as a mean free path for emitted photons, which depends on the imaginary part of the permittivity of the bulk scintillator as well as scattering-related losses. Due to the non-equilibrium nature of the scintillation process, it is possible that $\lambda$ generally depends on the properties of the excitation (energy and bandwidth). We generally assume here that both $C(E)$ and $\lambda$ are material- and energy-dependent.

We experimentally observed remote outcoupling across two bulk scintillators emitting at different wavelengths (YAG:Ce, 550~\unit{\nano\meter} and YAG:Yb, 1030~\unit{\nano\meter}), four orders of magnitude of excitation energy ($10^0$--$10^4$~\unit{\eV}), and monochromatic and broadband sources. Using our experimental setup illustrated in Fig.~\ref{fig:light_collecting}(a), we captured images under different source and slit conditions. From these images, we extracted $2a$, $y_0$, and the average pattern intensity, $\overline{I}_i$ (where $i$ is the pattern number), for each pattern in the sample. When multiple patterns were available for a given combination of parameters $(a,y_0,E)$, $\overline{I}_i$-values were averaged across all the patterns to consolidate the data, resulting in $\overline{I}_{p}(a,y_0,E)$ (where $E$ is the excitation energy). Each $(a,y_0,E)$ dataset for each scintillator was normalized such that the maximum value of $\overline{I}_i$ is equal to 1. The data are fitted according to Eq. (\ref{eq:dipole_model}). In Fig.~\ref{fig:light_collecting}(c), we plot the results of our analysis (detailed in Supplementary Material S3C). Experimental data points $\overline{I}_{p}(a,y_0,E)$ and the fitted curves $I_E(a,y_0)$ are indicated by the markers and solid lines, respectively. Error bars are shown for $\overline{I}_{p}(a,y_0,E)$, while the shaded areas surrounding the solid lines indicate the 95\% confidence interval.

\section*{Discussion}

In this work, we demonstrated volumetric nanophotonic patterning in bulk scintillators with nearly subwavelength feature sizes and hundreds of microns in depth, using direct femtosecond laser writing. Our 2.5D, volumetric nanophotonic scintillators follow expected trends observed for 2D photonic crystals. In particular, scintillation enhancement increases as the periods of the patterns approach the subwavelength regime~\cite{Yu2010pnas, RoquesCarmes2022}. The observation of these established trends suggests that photonic designs from other application areas such as thermal photonics can be extended to volumetric nanophotonic scintillators. We also observed an increase in scintillation enhancement as a function of etch depth. Furthermore, we discovered a ``remote outcoupling'' effect in which patterns produce a signal despite not being directly excited. We demonstrated the robustness of this effect across different bulk scintillators, for different emission energies ranging from green to near-IR light, as well as for different sources with energies spanning four orders of magnitude across different experiments (UV, $10^0$~\unit{\eV}-scale photons to X-ray, $10^4$~\unit{\eV}-scale photons). To explain this phenomenon, we developed a simple 2D model for laterally-propagating light that is scattered towards the camera by the nanophotonic patterns, with good agreement when fitted to our data.

We believe that our results are an important step toward total control over scintillation using nanophotonics, with the potential to integrate 3D nanophotonic effects such as photonic band gaps and Purcell enhancement into scintillators. Further volumetric effects to explore are, for example, the possibility for coupling between stopping power and light extraction, since the attenuation of high-energy particles and their conversion into lower-energy photons is depth-dependent~\cite{Beer2025,Choi2026b}. Finally, we believe that remote light collection has the potential to be a useful functionality of volumetric nanophotonic scintillators, enabling the detection of high-energy particles at a remote distance with line of sight from their point of impact. Our observation of remote outcoupling also highlights that quantitative predictions of scintillation enhancement must account for additional sources of scattering beyond the reciprocal absorption enhancement captured by idealized simulations. These include scattering from the surface and patterning roughness, fabrication-induced aperiodicity, and light generated outside the patterned region. Overall, our work points toward new photonic design strategies for controlling scintillation in the bulk, with potential applications in medical imaging and particle detection.

\section*{Methods}

\subsection{Fabrication}
The samples investigated in this work are YAG:Ce and YAG:Yb inorganic crystals. Their nanofabrication involves three steps: femtosecond laser writing, mechanical grinding and polishing, and chemical wet etching.

\textbf{Femtosecond laser writing.} In the first step, we employ direct femtosecond laser writing  to locally modify the crystal structure and induce a large etching selectivity between the pristine and photo-irradiated regions of the YAG host crystal. In undoped YAG, etching selectivity exceeding $10^{5}$ has been demonstrated \cite{Rodenas2019}, corresponding to etching rates of approximately \qty{100}{\micro\metre} per hour. Laser inscription is performed using a Menlo Systems Bluecut laser operating at a wavelength of \qty{1030}{\nano\meter}, with a pulse duration of approximately \qty{260}{\femto\second}, a pulse repetition rate of \qty{500}{\kilo\hertz}, and pulse energies in the range of \qty{160}{\nano\joule} to \qty{350}{\nano\joule}. The beam is focused into the sample using a single lens ($f$~=~\qty{3.1}{\milli\meter}, NA~=~0.68; Thorlabs C330TMD-B), and the structures are written at a translation speed of \qty{1}{\milli\meter\per\second}. The laser exposure begins at the bottom of the crystal and proceeds upward by translating the sample downwards relative to the focus. To prevent the formation of craters ($>$\qty{2}{\micro\meter} in diameter) caused by surface ablation, the laser is switched off approximately \qtyrange{30}{50}{\micro\meter} below the top surface using the internal acousto-optic modulator (AOM), ensuring that the modified regions remain embedded within the bulk material. To compensate for depth-dependent optical aberrations---which reduce peak intensity at the focal point---we implement a linear power ramp during the fabrication process. We start  with \qty{350}{\nano\joule} at the bottom surface and decrease the writing power to \qty{160}{\nano\joule} at the top surface. The laser power is chosen to remain just above the threshold required for chemical etching, while avoiding higher powers that would enlarge the modified (and thus etchable) regions and furthermore lead to the formation of cracks in the sample.

\textbf{Mechanical grinding and polishing.} Following laser inscription, the laser-modified regions are exposed by mechanical grinding and polishing using an Allied High Tech Products MultiPrep System operated at 100 rpm. Material is first removed from the bottom surface (typically a few hundred microns) using a \qty{30}{\micro\meter} diamond abrasive paper. Subsequent steps employ \qty{10}{\micro\meter} and \qty{3}{\micro\meter} diamond abrasive papers, removing approximately \qtyrange{30}{50}{\micro\meter} and \qty{10}{\micro\meter}, respectively. On the top surface, the coarsest grinding step is omitted.

\textbf{Chemical wet etching.} In the final step, the samples are chemically wet-etched. Prior to etching, the samples are cleaned in acetone and isopropanol ultrasonic baths (\qty{5}{\minute} each), followed by rinsing in deionized water. The etching is performed in 44 wt\% phosphoric acid ($\mathrm{H_3PO_4}$) at \qty{78}{\celsius} on a hotplate. The sample is held vertically using a custom-made Teflon holder, while the solution is stirred using a magnetic stirrer and covered with an hourglass to minimize evaporation. After etching, the cleaning procedure is repeated. Typical etching times are several hours depending on the desired etching depth; the process can be repeated for deeper etching depths.
Further details on the fabrication procedure can be found in Ref.~\cite{Jurgensen2024}.

\subsection{General experimental setup}
In this section, we describe our experimental setup in broad strokes; much more detailed descriptions and figures can be found in Supplementary Materials S1. All of the experimental setups consist of (1) a source, (2) preparation optics, (3) a scintillator, (4) collection optics, including a filter to reduce background, and (5) a camera. The details of these general components depended on the experiment being conducted, as discussed below.

\subsubsection{UV laser setup}
In these experiments, the source was a 349~\unit{\nano\meter} (or 3.55~\unit{\eV}) UV laser, which was fiber-coupled to a collimator pointed at a slit (preparation optics). A mechanical movement was used to continuously vibrate the fiber to reduce speckle. The scintillator was mounted to an XY translation stage, which itself was screwed into a Z translation stage for fine focusing. For the image shown in Fig.~\ref{fig:air_holes}(b)(ii), we used a 50$\times$ microscope objective (Nikon L Plan, NA = 0.45), while for data shown in Fig.~\ref{fig:light_collecting}(c), we used a 5$\times$ microscope objective (Mitutoyo 46-402, NA = 0.14). Both microscope objectives were coupled to a tube lens, with a narrowband filter centered at or near the emission wavelength of the scintillator placed between them (Semrock FF01-560/14-25). The camera was a Hamamatsu C14440-20UP.

In Fig.~\ref{fig:light_collecting}, we only report light collection under UV photon excitation for YAG:Ce. The reason is that the signal from YAG:Yb was too weak to be detected by the camera as a result of the compounding effects of low excitation energy (compared to X-ray photons, which are nearly four orders of magnitude more energetic) and $\lesssim5$\% quantum efficiency of the camera at 1030~\unit{\nano\meter}.

\subsubsection{Synchrotron setup (monochromatic X-ray excitation)}
Synchrotron experiments were conducted on the B16 beamline at the Diamond Light Source (United Kingdom). The beam energy was controlled using a double crystal monochromator; measurements were recorded at energies of 8, 11, 14, 17, 20, 23, or 25~keV for YAG:Yb and 8, 11.8, 17, 20.1, or 22.2~keV for YAG:Ce. Two adjustable tungsten slits (JJ X-Ray) were used to control the size of the beam. The scintillator was mounted to a stack of piezo stages (attocube), enabling precise XYZ translation and rotation about three axes. To image the scintillator, a microscope objective with 10$\times$ magnification and an NA of 0.4 was used (Optique Peter). Bandpass filters were used to measure scintillation at the peak wavelength of each sample (Semrock FF01-550/200-25 for YAG:Ce and Thorlabs FLH1030-10 for YAG:Yb). Images were captured using a Hamamatsu ORCA-Quest qCMOS camera.

\subsubsection{Micro-CT setup (broadband X-ray excitation)}
The broadband X-ray source we used was part of a commercially-available ZEISS Xradia 620 Versa micro-CT. The sample holder, which was different for scintillation enhancement and light collection experiments, was placed on a 4-axis sample stage (XYZ translation and rotation about Z). For the scintillation enhancement experiments, the sample holder was a clamp that came with the micro-CT. For the light collection experiments, the sample holder consisted of a slit and an XY translation stage mounted on an optical cage system (the same as the UV laser setup). The samples were imaged using either a 50$\times$ (Nikon L Plan, NA = 0.45) or 10$\times$ (Nikon Plan Fluor, NA = 0.30) microscope objective directly mounted on the turret of the micro-CT, which is coupled to an Andor iKon-L DW936N-BV camera. The filter (Semrock FF01-560/14-25 for YAG:Ce and Thorlabs FLH1030-10 for YAG:Yb) was taped to the back of the microscope objective before mounting to the turret. In the main text as well as Supplementary Materials S1B and S3B, we describe our strategy for managing low signal from YAG:Yb and subsequent etaloning.

\subsection{Theory of remote outcoupling}
A full derivation of our simple model of remote light outcoupling can be found in Supplementary Materials S5. Here, we generally describe our approach, starting with the following assumptions: (1) the problem is essentially 2D; (2) the slit or excitation region is effectively infinite along one direction (call it $x$, which is perpendicular to $y$); (3) a uniform distribution of isotropically-emitting dipoles is generated in the excitation region by the impacting high-energy particles; and (4) the patterns are point objects located at $(x_0,y_0)$. We assume the radiated power emerging from the dipoles spreads cylindrically, scaling as $1/r$ with an exponential attenuation with $r$ to incorporate loss. Therefore, we argue that the propagation kernel of emitted light is given by $g(r)=e^{-r/\lambda}/(2\pi r)$; here, $\lambda$ represents the mean free path of photons. By considering the intensity contribution of a differential element $dA=dx\,dy$ in the excitation region, we can show that the intensity of light collected by the pattern is:
\begin{widetext}
\begin{equation}
    I_E(a,y_0)=\frac{\varepsilon P_{0}\sigma(E)}{2\pi}\int_{-a}^{a}\int_{-\infty}^{\infty}\frac{e^{-\left(\sqrt{(x-x_0)^2+(y-y_0)^2}\right)/\lambda}}{\sqrt{(x-x_0)^2+(y-y_0)^2}}dx\,dy,
\end{equation}
\end{widetext}
where $\varepsilon\in[0,1]$ is the collection efficiency, $P_0$ is the emitted power by each dipole, and $\sigma(E)$ is the density of dipoles. Evaluating this integral results in Eq. (\ref{eq:dipole_model}).

\section*{Acknowledgments}
We thank Christopher Hogan from the Institute for Soldier Nanotechnologies at MIT for help with operating and troubleshooting the experimental tools used in this work. We thank Shiekh Zia Uddin for help with building the speckle-reducing device used in our UV laser experiments. S. Pajovic gratefully acknowledges support from the NSF GRFP under Grant No. 2141064 as well as the MathWorks Engineering Fellowship via MIT. J. Chen and W. Michaels also gratefully acknowledge support from the NSF GRFP under Grant No. 2637797. C.R.C. was supported by a Stanford Science Fellowship and acknowledges startup funding from the Institute of Science and Technology Austria (ISTA). Work completed at the STFC Rutherford Appleton Laboratory was funded by the UKRI STFC Centre for Instrumentation Engineered Scintillators programme, 2023--2026. We thank Diamond Light Source for time on beamline B16 (proposal numbers OM35423 and OM40503) that contributed to the results presented here, and the assistance of the beamline staff, particularly Vishal Dhamgaye (beamline scientist), Andrew Malandain (beamline technician), Andrew Male, and Ali Anari (design and manufacture of quick release sample mounts). This material is based upon work supported in part by the U.S. Army Research Office through the Institute for Soldier Nanotechnologies at MIT, made possible by the U.S. Army DEVCOM ARL Army Research Office, under Cooperative Agreement number W911NF-23-2-0121. We gratefully acknowledge the support of the ONR and ARO MURI programs under agreement numbers N00014-20-1-2325 and W911NF-24-1-0224, respectively. This work was also supported in part by the Army Research Office under MURI Grant No. W911NF-26-1-A293. We also acknowledge support from the ONR under grant number N00014-23-1-2102. We acknowledge partial support from the U.S. National Science Foundation under award number 2614673.

\nocite{*}

\bibliography{references}% Produces the bibliography via BibTeX.

@article{Rodenas2019,
	title     = {{Three-dimensional femtosecond laser nanolithography of crystals}},
	author    = {Ródenas, Airán and Gu, Min and Corrielli, Giacomo and Paiè, Petra and John, Sajeev and Kar, Ajoy K. and Osellame, Roberto},
	year      = 2019,
	journal   = {Nature Photonics},
	volume    = 13,
	pages     = {105--109}
}

@phdthesis{Jurgensen2024,
	title     = {{Emergent Quantization in Interacting Thouless Pumps}},
	author    = {Jurgensen, Marius},
	year      = 2024,
	school    = {{The Pennsylvania State University}}
}

@article{RoquesCarmes2022,
	title     = {{A framework for scintillation in nanophotonics}},
	author    = {Roques-Carmes, Charles and Rivera, Nicholas and Ghorashi, Ali and Kooi, Steven E. and Yang, Yi and Lin, Zin and Beroz, Justin and Massuda, Aviram and Sloan, Jamison and Romeo, Nicolas and Yu, Yang and Joannopoulos, John D. and Kaminer, Ido and Johnson, Steven G. and Soljačić, Marin},
	year      = 2022,
	journal   = {Science},
	volume    = 375,
	number    = 6583,
	pages     = {eabm9293}
}

@article{MartinMonier2024,
	title     = {{Large-scale self-assembled nanophotonic scintillators for X-ray imaging}},
	author    = {Martin-Monier, Louis and Pajovic, Simo and Abebe, Muluneh G. and Chen, Joshua and Vaidya, Sachin and Min, Seokhwan and Choi, Seou and Kooi, Steven E. and Maes, Bjorn and Hu, Juejun and Soljačić, Marin and Roques-Carmes, Charles},
	year      = 2025,
	journal   = {Nature Communications},
	volume    = 16,
	pages     = 5750
}

@article{Yu2010pnas,
	title     = {{Fundamental limit of nanophotonic light trapping in solar cells}},
	author    = {Yu, Zongfu and Raman, Aaswath and Fan, Shanhui},
	year      = 2010,
	journal   = {Proceedings of the National Academy of Sciences},
	volume    = 107,
	number    = 41,
	pages     = {17491--17496}
}

@misc{etaloning,
	title     = {{Etaloning in Back Illuminated CCDs}},
	author    = {{Roper Scientific, Inc.}},
	year      = 2000,
	note      = {Technical note}
}

@article{Necas2012,
	title     = {{Gwyddion: an open-source software for {SPM} data analysis}},
	author    = {Nečas, David and Klapetek, Petr},
	year      = 2012,
	journal   = {Central European Journal of Physics},
	volume    = 10,
	pages     = {181--188}
}

@article{ruptures,
	title     = {{Selective review of offline change point detection methods}},
	author    = {Truong, Charles and Oudre, Laurent and Vayatis, Nicolas},
	year      = 2020,
	journal   = {Signal Processing},
	volume    = 167,
	pages     = 107299
}

@book{Gradshteyn2014,
	title     = {{Table of Integrals, Series, and Products}},
	author    = {Gradshteyn, Izrail Solomonovich and Ryzhik, Iosif Moiseevich},
	year      = 2014,
	publisher = {Academic Press},
	editor    = {Zwillinger, Daniel and Moll, Victor},
	edition   = 8
}

@article{Wang2023,
	title     = {{Needs, Trends, and Advances in Scintillators for Radiographic Imaging and Tomography}},
	author    = {Wang, Zhehui and Dujardin, Christophe and Freeman, Matthew S. and Gehring, Amanda E. and Hunter, James F. and Lecoq, Paul and Liu, Wei and Melcher, Charles L. and Morris, Christopher L. and Nikl, Martin and Pilania, Ghanshyam and Pokharel, Reeju and Robertson, Daniel G. and Rutstrom, Daniel J. and Sjue, Sky K. and Tremsin, Anton S. and Watson, Scott A. and Wiggins, Brenden W. and Winch, Nicola M. and Zhuravleva, Mariya},
	year      = 2023,
	journal   = {IEEE Transactions on Nuclear Science},
	volume    = 70,
	number    = 7,
	pages     = {1287--1316}
}

@article{Singh2024,
	title     = {{Bright Innovations: Review of Next-Generation Advances in Scintillator Engineering}},
	author    = {Singh, Pallavi and Dosovitskiy, Georgy and Bekenstein, Yehonadav},
	year      = 2024,
	journal   = {ACS Nano},
	volume    = 18,
	number    = 22,
	pages     = {14029--14049}
}

@article{Kurman2020,
	title     = {{Photonic-Crystal Scintillators: Molding the Flow of Light to Enhance X-Ray and {$\gamma$}-Ray Detection}},
	author    = {Kurman, Yaniv and Shultzman, Avner and Segal, Ohad and Pick, Adi and Kaminer, Ido},
	year      = 2020,
	journal   = {Physical Review Letters},
	volume    = 125,
	number    = 4,
	pages     = {040801}
}

@article{Lin2022,
	title     = {{Structured Scintillators for Efficient Radiation Detection}},
	author    = {Lin, Ziyu and Lv, Shichao and Yang, Zhongmin and Qiu, Jianrong and Zhou, Shifeng},
	year      = 2022,
	journal   = {Advanced Science},
	volume    = 9,
	number    = 2,
	pages     = 2102439
}

@article{Sabet2016,
	title     = {{Novel Laser-Processed CsI:Tl Detector for SPECT}},
	author    = {Sabet, Habib and Bl{\"a}ckberg, Lars and Uzun-Ozsahin, Didem and El Fakhri, Georges},
	year      = 2016,
	journal   = {Medical Physics},
	volume    = 43,
	number    = 5,
	pages     = {2630--2638}
}

@article{Ye2022,
	title     = {{Enhancing Large-Area Scintillator Detection with Photonic Crystal Cavities}},
	author    = {Ye, Wenzheng and Bizarri, Gregory and Birowosuto, Muhammad Danang and Wong, Liang Jie},
	year      = 2022,
	journal   = {ACS Photonics},
	volume    = 9,
	number    = 12,
	pages     = {3917--3925}
}

@article{Makowski2025,
	title     = {{Scaling Up Purcell-Enhanced Self-Assembled Nanoplasmonic Perovskite Scintillators into the Bulk Regime}},
	author    = {Makowski, Micha{\l} and Ye, Wenzheng and Kowal, Dominik and Maddalena, Francesco and Mahato, Somnath and Amrillah, Yudhistira Tirtayasri and Zajac, Weronika and Witkowski, Marcin Eugeniusz and Drozdowski, Konrad Jacek and Nathaniel and Dang, Cuong and Cybinska, Joanna and Drozdowski, Winicjusz and Nugroho, Ferry Anggoro Ardy and Dujardin, Christophe and Wong, Liang Jie and Birowosuto, Muhammad Danang},
	year      = 2025,
	journal   = {Advanced Materials},
	volume    = 37,
	number    = 30,
	pages     = {e2417874}
}

@book{Lecoq2017,
	title     = {{Inorganic Scintillators for Detector Systems: Physical Principles and Crystal Engineering}},
	author    = {Lecoq, Paul and Gektin, Alexander and Korzhik, Mikhail},
	year      = 2017,
	publisher = {Springer Cham},
	edition   = 2
}

@article{Kurman2024,
	title     = {{Purcell-enhanced X-ray scintillation}},
	author    = {Kurman, Yaniv and Lahav, Neta and Schuetz, Roman and Shultzman, Avner and Roques-Carmes, Charles and Lifshits, Alon and Zaken, Segev and Lenkiewicz, Tom and Strassberg, Rotem and Be'er, Orr and {others}},
	year      = 2024,
	journal   = {Science Advances},
	volume    = 10,
	number    = 44,
	pages     = {eadq6325}
}

@article{Long2024,
	title     = {{Nonreciprocal scintillation using one-dimensional magneto-optical photonic crystals}},
	author    = {Long, Olivia and Pajovic, Simo and Roques-Carmes, Charles and Tsurimaki, Yoichiro and Rivera, Nicholas and Soljacic, Marin and Boriskina, Svetlana and Fan, Shanhui},
	year      = 2024,
	journal   = {Physical Review Applied},
	volume    = 22,
	number    = 5,
	pages     = {054062}
}

@article{Shultzman2024,
	title     = {{Toward a Second Generation of Metascintillators Using the Purcell Effect}},
	author    = {Shultzman, Avner and Sch{\"u}tz, Roman and Kurman, Yaniv and Lahav, Neta and Dosovitskiy, George and Roques-Carmes, Charles and Bekenstein, Yehonadav and Konstantinou, Georgios and Latella, Riccardo and Zhang, Lei and others},
	year      = 2024,
	journal   = {IEEE Transactions on Radiation and Plasma Medical Sciences},
	volume    = 9,
	number    = 2,
	pages     = {141--147}
}

@article{Shultzman2023,
	title     = {{Enhanced Imaging Using Inverse Design of Nanophotonic Scintillators}},
	author    = {Shultzman, Avner and Segal, Ohad and Kurman, Yaniv and Roques-Carmes, Charles and Kaminer, Ido},
	year      = 2023,
	journal   = {Advanced Optical Materials},
	volume    = 11,
	number    = 8,
	pages     = 2202318
}

@article{RoquesCarmes2023,
	title     = {{Free-electron--light interactions in nanophotonics}},
	author    = {Roques-Carmes, Charles and Kooi, Steven E and Yang, Yi and Rivera, Nicholas and Keathley, Phillip D and Joannopoulos, John D and Johnson, Steven G and Kaminer, Ido and Berggren, Karl K and Solja{\v{c}}i{\'c}, Marin},
	year      = 2023,
	journal   = {Applied Physics Reviews},
	volume    = 10,
	number    = 1,
	pages     = {011303}
}

@article{Gattass2008,
	title     = {{Femtosecond laser micromachining in transparent materials}},
	author    = {Gattass, Rafael R. and Mazur, Eric},
	year      = 2008,
	journal   = {Nature Photonics},
	volume    = 2,
	pages     = {219--225}
}

@article{Sabet2024,
	title     = {{Laser nanofabrication inside silicon with spatial beam modulation and anisotropic seeding}},
	author    = {Asgari Sabet, Rana and Ishraq, Aqiq and Saltik, Alperen and B{\"u}t{\"u}n, Mehmet and others},
	year      = 2024,
	journal   = {Nature Communications},
	volume    = 15,
	pages     = 5786
}

@article{PazBuclatin2026,
	title     = {{Fabrication of a monolithic all-YAG crystalline microresonator through femtosecond laser nanolithography and thermal annealing}},
	author    = {Paz-Buclatin, Franzette and Martin, Leopoldo Luis and Gonz{\'a}lez-Orive, Alejandro and Gonz{\'a}lez-Tombolato, Urma and Kamada, Kei and Yoshikawa, Akira and Segui, Airan Rodenas},
	year      = 2026,
	journal   = {Optics and Lasers in Engineering},
	volume    = 199,
	pages     = 109562
}

@article{PazBuclatin2025,
	title     = {{Three-dimensional nanofabrication of hollow wavelength-thick lenses deep inside solid-state optical crystals}},
	author    = {Paz-Buclatin, Franzette and Molina, Pablo and Gonz{\'a}lez-Tombolato, Urma and Kamada, Kei and Yoshikawa, Akira and Martin, Leopoldo Luis and R{\'o}denas, Air{\'a}n},
	year      = 2025,
	journal   = {Optics \& Laser Technology},
	volume    = 192,
	pages     = 114034
}

@article{Han2022,
	title     = {{Three-dimensional nanofabrication via ultrafast laser patterning and kinetically regulated material assembly}},
	author    = {Han, Fei and Gu, Songyun and Klimas, Aleks and Zhao, Ni and Zhao, Yongxin and Chen, Shih-Chi},
	year      = 2022,
	journal   = {Science},
	volume    = 378,
	number    = 6626,
	pages     = {1325--1331}
}

@article{Oran2018,
	title     = {{3D nanofabrication by volumetric deposition and controlled shrinkage of patterned scaffolds}},
	author    = {Oran, Daniel and Rodriques, Samuel G and Gao, Ruixuan and Asano, Shoh and Skylar-Scott, Mark A and Chen, Fei and Tillberg, Paul W and Marblestone, Adam H and Boyden, Edward S},
	year      = 2018,
	journal   = {Science},
	volume    = 362,
	number    = 6420,
	pages     = {1281--1285}
}

@article{Zhang2024,
	title     = {{Printing of 3D photonic crystals in titania with complete bandgap across the visible spectrum}},
	author    = {Zhang, Wang and Min, Jiakang and Wang, Hao and Wang, Hongtao and Li, Xue Liang and Ha, Son Tung and Zhang, Biao and Pan, Cheng-Feng and Li, Hao and Liu, Hailong and Yin, Hui and Yang, Xiaolong and Liu, Siqi and Xu, Xiaodong and He, Chaobin and Yang, Hui Ying and Yang, Joel K. W.},
	year      = 2024,
	journal   = {Nature Nanotechnology},
	volume    = 19,
	number    = 12,
	pages     = {1813--1820}
}

@book{AbramowitzStegun,
	title     = {{Handbook of Mathematical Functions: with Formulas, Graphs, and Mathematical Tables}},
	year      = 1972,
	publisher = {United States Department of Commerce, National Bureau of Standards},
	editor    = {Abramowitz, Milton and Stegun, Irene A.},
	edition   = 10
}

@article{Zhu2015a,
  title={{Enhanced light extraction of scintillator using large-area photonic crystal structures fabricated by soft-X-ray interference lithography}},
  author={Zhu, Zhichao and Wu, Shuang and Xue, Chaofan and Zhao, Jun and Wang, Liansheng and Wu, Yanqing and Liu, Bo and Cheng, Chuanwei and Gu, Mu and Chen, Hong and others},
  journal={Applied Physics Letters},
  volume={106},
  number={24},
  year={2015},
  pages={241901},
}

@unpublished{Zanettini2016,
  title={{Improved Light Extraction Efficiency on 2 inches LYSO with Nanopatterned TiO$_2$ Photonic Crystals}},
  author={Zanettini, S and Gâté, V and Usureau, E and Ruscica, J and Hamouda, F and Nomenyo, K and Le Cunff, L and Kadiri, H and Lerondel, G and Salomoni, M and others},
  note={{IEEE conference, Strasbourg (2016)}}
}

@article{Chen2018,
  title={{Enhanced light extraction of plastic scintillator using large-area photonic crystal structures fabricated by hot embossing}},
  author={Chen, Xueye and Liu, Bo and Wu, Qiang and Zhu, Zhichao and Zhu, Jingtao and Gu, Mu and Chen, Hong and Liu, Jinliang and Chen, Liang and Ouyang, Xiaoping},
  journal={Optics Express},
  volume={26},
  number={9},
  pages={11438--11446},
  year={2018}
}

@article{Liu2017,
  title={{Modified timing characteristic of a scintillation detection system with photonic crystal structures}},
  author={Liu, Jinliang and Liu, Bo and Zhu, Zhichao and Chen, Liang and Hu, Jing and Xu, Mengxuan and Cheng, Chuanwei and Ouyang, Xiaoping and Zhang, Zhongbing and Ruan, Jinlu and others},
  journal={Optics Letters},
  volume={42},
  number={5},
  pages={987--990},
  year={2017}
}

@article{Zhang2017,
  title={{Enhanced light extraction of LYSO scintillator by photonic crystal structures from a modified porous anodized aluminum oxide layer}},
  author={Zhang, Juannan and Liu, Bo and Zhu, Zhichao and Wu, Qiang and Cheng, Chuanwei and Liu, Jinliang and Chen, Liang and Ouyang, Xiaoping and Gu, Mu and Xu, Jun and others},
  journal={Nuclear Instruments and Methods in Physics Research Section A: Accelerators, Spectrometers, Detectors and Associated Equipment},
  volume={864},
  pages={36--39},
  year={2017}
}

@article{Zhu2015b,
  title={{Improvement of light extraction of LYSO scintillator by using a combination of self-assembly of nanospheres and atomic layer deposition}},
  author={Zhu, Zhichao and Liu, Bo and Zhang, Haifeng and Ren, Weina and Cheng, Chuanwei and Wu, Shuang and Gu, Mu and Chen, Hong},
  journal={Optics Express},
  volume={23},
  number={6},
  pages={7085--7093},
  year={2015}
}

@article{Zhu2015c,
  title={{Improved light extraction of LYSO scintillator by the photonic structure from a layer of anodized aluminum oxide}},
  author={Zhu, Zhichao and Liu, Bo and Cheng, Chuanwei and Zhang, Haifeng and Wu, Shuang and Gu, Mu and Chen, Hong and Chen, Liang and Liu, Jinliang and Ouyang, Xiaoping},
  journal={Nuclear Instruments and Methods in Physics Research Section A: Accelerators, Spectrometers, Detectors and Associated Equipment},
  volume={786},
  pages={1--4},
  year={2015}
}

@article{Knapitsch2014,
  title={{Review on photonic crystal coatings for scintillators}},
  author={Knapitsch, Arno and Lecoq, Paul},
  journal={International Journal of Modern Physics A},
  volume={29},
  number={30},
  pages={1430070},
  year={2014}
}

@article{Salomoni2018,
  title={{Photonic Crystal Slabs Applied to Inorganic Scintillators}},
  author={Salomoni, Matteo and Pots, Rosalinde and Lecoq, Paul and Auffray, Etiennette and Gundacker, Stefan and Paganoni, Marco and Singh, Bipin and Marshall, Matthew and Nagarkar, Vivek V},
  journal={IEEE Transactions on Nuclear Science},
  volume={65},
  number={8},
  pages={2191--2195},
  year={2018}
}

@misc{Zabel2026,
  title={{Dosimetric characterization of a nanophotonic scintillator and applications to real-time in-vivo total body irradiation dosimetry}},
  author={Zabel, W Jeffrey and Chen, Dixin and Martin-Monier, Louis and Pajovic, Simo and Fan, Shanhui and Hu, Juejun and Solja{\v{c}}i{\'c}, Marin and Xing, Lei and Roques-Carmes, Charles and Ashraf, M Ramish},
  year = {2026},
  note={arXiv:2606.16007}
}

@article{Wollesen2022,
  title={{Scintillating thin film design for ultimate high resolution X-ray imaging}},
  author={Wollesen, Laura and Riva, Federica and Douissard, Paul-Antoine and Pauwels, Kristof and Martin, Thierry and Dujardin, Christophe},
  journal={Journal of Materials Chemistry C},
  volume={10},
  number={24},
  pages={9257--9265},
  year={2022}
}

@misc{Choi2026a,
  title={Supercollimating photonic crystal scintillators},
  author={Choi, Seou and Vaidya, Sachin and Roques-Carmes, Charles and Solja{\v{c}}i{\'c}, Marin},
  year = {2026},
  note={arXiv:2605.17006}
}

@article{Beer2025,
  title={{Heterostructure Nanoscintillator for Matching Radiation Absorbing Layers with Fast Light-Emitting Layers}},
  author={Be’er, Orr and Shultzman, Avner and Strassberg, Rotem and Dosovitskiy, Georgy and Veber, Noam and Schuetz, Roman and Roques-Carmes, Charles and Kaminer, Ido and Bekenstein, Yehonadav},
  journal={Nano Letters},
  volume={25},
  number={9},
  pages={3422},
  year={2025}
}

@misc{Choi2026b,
  title={Hybrid Nanophotonic Scintillators for Enhanced X-ray Absorption, Emission, and Time Resolution},
  author={Choi, Seou and Vaidya, Sachin and Shultzman, Avner and Roques-Carmes, Charles and Kaminer, Ido and Solja{\v{c}}i{\'c}, Marin},
  year = {2026},
  note={arXiv:2605.14992}
}

@misc{NISThandbook,
  title={{NIST/SEMATECH e-Handbook of Statistical Methods}},
  year={2003},
  note={Ch. 4.5.1.1, https://itl.nist.gov/div898/handbook/pmd/section5/pmd511.htm}
}

@inbook{SeberWild,
    author={Seber, G. A. and Wild, C. J.},
    title={{Nonlinear Regression}},
    chapter={5.1},
    pages={191--194},
    publisher={Wiley},
    year={1989},
    address={New York, NY}
}

@inbook{Robinson2016,
    author={Robinson, Edward L.},
    title={{Data Analysis for Scientists and Engineers}},
    chapter={4.3},
    pages={86--91},
    publisher={Princeton University Press},
    year={2016},
    address={Princeton, NJ}
}

\end{document}

\rmfamily

\title{Supplementary Materials for: Volumetric Nanophotonics for Enhanced Scintillation and Remote Light Outcoupling}

\author{Simo~Pajovic$^{\bigstar,1,\ddagger}$}
\email{pajovics@mit.edu}
\author{Marius~Jürgensen$^{\bigstar,2,3}$}
\author{Sachin~Vaidya$^{4, 5}$}
% \email{svaidya1@mit.edu}
\author{Johnathon~P.~Gales$^{3}$}
\author{Joshua~Chen$^{5}$}
\author{Steven~E.~Kooi$^{6}$}
\author{Sion~Richards$^{7}$}
\author{Isabel~H.~B.~Braddock$^{7}$} 
\author{Karen~Mamian$^{8}$}
\author{William~Michaels$^{5}$}
\author{Shaul~Katznelson$^{8,9}$}
\author{Chris~D.~Armstrong$^{7}$}
\author{Ido~Kaminer$^{9}$}
\author{Marin~Soljačić$^{4, 5}$}
\author{Mikael~C.~Rechtsman$^{3}$}
\author{Charles~Roques-Carmes$^{5,8,10}$}
\email{crc@ista.ac.at}

\affiliation{\small $^{1}$Department of Mechanical Engineering, Massachusetts Institute of Technology, Cambridge, MA, USA\looseness=0}
\affiliation{\small $^{2}$Department of Physics, Stanford University, Stanford, CA, USA\looseness=0}
\affiliation{\small $^{3}$Department of Physics, The Pennsylvania State University, University Park, PA, USA\looseness=0}
\affiliation{\small $^{4}$Department of Physics, Massachusetts Institute of Technology, Cambridge, MA, USA\looseness=0}
\affiliation{\small $^{5}$Research Laboratory of Electronics, Massachusetts Institute of Technology, Cambridge MA USA\looseness=0}
\affiliation{\small $^{6}$Institute for Soldier Nanotechnologies, Massachusetts Institute of Technology, Cambridge, MA, USA\looseness=0}
\affiliation{\small $^{7}$Science and Technology Facilities Council, Rutherford Appleton Laboratory, Didcot, UK\looseness=0}
\affiliation{\small $^{8}$Institute of Science and Technology Austria (ISTA), Klosterneuburg, Austria\looseness=0}
\affiliation{\small $^{9}$Solid State Institute, Technion - Israel Institute of Technology, Haifa, Israel\looseness=0}
\affiliation{\small $^{10}$E. L. Ginzton Laboratory, Stanford University, Stanford, CA, USA\looseness=0}
\affiliation{\small $^{\ddagger}$Current address: Department of Applied Physics and Materials Science, California Institute of Technology, Pasadena, CA, USA\looseness=0\\
$^{\bigstar}$ denotes equal contribution.}

% \tableofcontents 

\renewcommand{\sp}{\sigma_+}
\newcommand{\sm}{\sigma_-}

%-------------------------------------
%------------- MAIN TEXT -------------
%-------------------------------------

\maketitle

%-----CHANGE SETUP FOR PARAGRAPH INDENTS AND SKIPS-----
\setlength{\parindent}{0em}
\setlength{\parskip}{.5em}
% \vspace*{-2em}

\tableofcontents

\newpage

\section{Details of the Experiments}\label{sec:experiments}

In this section, we further detail the three experimental setups used to characterize light emitted by the volumetric nanophotonic scintillators. All the experiments we performed were essentially the same, consisting of (1) a source to excite the sample, (2) optics to control the source, which depend on the nature of the source, (3) the sample, which scintillates, (4) optics to collect the scintillation, typically an infinity-corrected optical system (microscope objective and tube lens) in tandem with a filter to reduce background, and (5) a camera connected to a computer for data acquisition. For YAG:Yb (emission peak at 1030~\unit{\nano\meter}) we used a Thorlabs FLH1030-10 filter (center wavelength 1030~\unit{\nano\meter}, FWHM\footnote{Full width at half maximum.} 10~\unit{\nano\meter}); for YAG:Ce (emission peak at 550~\unit{\nano\meter}), we used an FF01-560/14-25 filter (center wavelength 560~\unit{\nano\meter}, FWHM 20~\unit{\nano\meter}) or an FF01-550/200-25 (center wavelength 550~\unit{\nano\meter}, FWHM 210.3~\unit{\nano\meter}) from Semrock. In what follows, we summarize the remaining details for each of the three sources and their experiments.

\subsection{UV Laser Setup}\label{subsec:uv}

\begin{figure}
    \centering
    \includegraphics[scale=1.5]{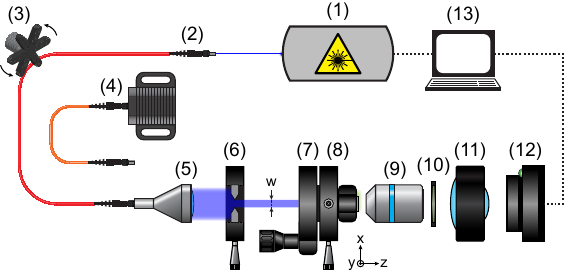}
    \caption{\textbf{Schematic of the UV laser setup and its components.} (1) Laser source. (2) Fiber coupler. (3) Spoked fiber vibration device (to reduce speckle). (4) Optional LED source. (5) Collimator. (6) Adjustable slit. (7) Z translation stage attached to (8) XY translation stage (axes shown). (9) Microscope objective. (10) Filter. (11) Tube lens. (12) Camera. (13) Computer controller. Components (5)--(11) were mounted on an optical cage system.}
    \label{suppfig:uv_laser}
\end{figure}

\textit{Setup:} The cage-mounted experimental setup, shown in Fig. \ref{suppfig:uv_laser}, includes a computer-controlled laser with peak emission at 349~\unit{\nano\meter} (Spectra-Physics Explorer One 349-120) fiber-coupled to a collimator (Thorlabs F810SMA-405). The pump current of the laser was always set to 1.5~\unit{\ampere}. The fiber connecting the laser to the collimator was placed in a device consisting of rotating spokes driven by a motor and microcontroller. The spokes continually vibrated the fiber as they rotated, reducing speckle. The fiber could be connected to a cold white LED (Thorlabs MCWHF2) as well, used for optical alignment. Laser light exiting the collimator passed through an adjustable slit (Thorlabs VA100CP) before impinging on the sample, which was mounted to an XY translation stage (Thorlabs CXY1QA). The sample itself was taped to a 3D-printed sample holder sandwiched between a pair of retaining rings. By using the collimator and placing the sample as close as possible to the slit, the effects of geometric magnification were minimized. Scintillation from the sample was collected by an infinity-corrected optical system, consisting of a microscope objective and a tube lens (Thorlabs TTL180-A). For the slit experiments, a $5\times$ microscope objective was used (Mitutoyo 46-402, WD\footnote{Working distance} = 37.5~\unit{\milli\meter}, NA\footnote{Numerical aperture.}=0.14), while for imaging, a $50\times$ one was used (Nikon L Plan, WD = 17~\unit{\milli\meter}, NA = 0.45). A filter was placed behind the microscope objective. Since the $50\times$ microscope objective had a shallow depth of field, when in use, the XY translation stage was also attached to a Z translation stage (Thorlabs SM1ZA) to facilitate focusing the image. Finally, the image was formed on a camera (Hamamatsu C14440-20UP) and saved using the included software (HCImage Live).

\textit{Slit experiments:} The goal of the slit experiments was to measure the scintillation from the patterns when they were not directly excited by the source, i.e., due to the slit. The dependent variables in this experiment were the slit width $w=2a$ and the center-to-center distances between the slit and the patterns $y_i$, where $i$ is used to denote different patterns on the same sample. The slit and the sample (more specifically, the patterns) were coarsely aligned and focused by hand under LED illumination and finalized under laser illumination. The initial position of the slit was always halfway between two rows of patterns to get as many data points as possible.

In the first set of experiments, we fixed $y_i$ and did a fine parameter sweep as a function of $w$. For each $w$-value, we recorded the micrometer reading of the slit and captured an image (exposure time 250~\unit{\milli\second}). This was repeated until we had collected $>20$ data points.

In the second set of experiments, we did a coarse parameter sweep as a function of both $y_i$ and $w$. For each $w$-value, we returned the slit to its initial position and recorded its micrometer reading. Then, using the XY translation stage, we manually moved the sample relative to the slit and captured images at several different $y_i$-values. Since the XY translation stage did not have a micrometer, the sample was moved by eye, and the $y_{i}$-value was estimated \textit{post hoc} from the images themselves (see Sec. \ref{sec:data_processing}). It was essential that the slit did not overlap with any of the samples so as not to directly excite them. This implied that for a given $w$-value, $y_{i} \geq w/2 + L_{i}/2$, where $L_{i}$ is the side length of pattern $i$. This constrained the number of $y_i$-values we could sweep over at larger $w$-values. At the end of each set of experiments, we captured a dark frame.

\subsection{Micro-CT Setup}\label{subsec:ct}

\begin{figure}
    \centering
    \includegraphics[scale=1.5]{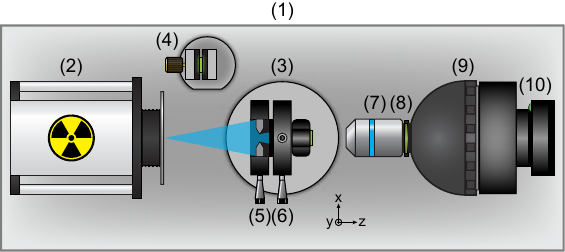}
    \caption{\textbf{Schematic of the micro-CT setup and its components.} (1) Enclosure. (2) X-ray source. (3) 4-axis sample stage. (4) Sample holder included with the micro-CT, used in the enhancement experiment. (5) Adjustable slit. (6) XY translation stage (axes shown). (5) and (6) were mounted on an optical cage system, then on the sample stage and used in the slit experiment. (7) Microscope objective and (8) filter mounted on a turret. (9) Turret and optics. (10) Camera.}
    \label{suppfig:ct}
\end{figure}

\textit{Setup:} The micro-CT experiments were performed using a ZEISS Xradia 620 Versa. The setup, shown in Fig. \ref{suppfig:ct}, consists of a cone beam X-ray source; a 4-axis (XYZ and rotation about Z) sample stage; a microscope objective (a 10$\times$ Nikon Plan Fluor, WD = 16~\unit{\milli\meter}, NA = 0.30 or a 50$\times$ Nikon L Plan, WD = 17~\unit{\milli\meter}, NA = 0.45); a filter to reduce background; and a detector (Andor iKon-L DW936N-BV). The source, sample stage, and detector are computer-controlled. In the scintillation enhancement experiments, the sample was mounted to a sample holder that came with the micro-CT. In the slit experiments, we used a custom-made and designed adapter~\cite{MartinMonier2024} made of aluminum (machining services provided by Protolabs) to mount an optical post to the sample stage, with a cage system similar to the UV laser setup. The cage system included the slit, an XY translation stage to control the position of the slit relative to the patterns on the sample, and a 3D-printed sample holder.

\textit{Enhancement experiment:} To measure the scintillation enhancement of the patterns, we set the voltage and power of the source to 160 kVp and 25~\unit{\watt}, meaning the maximum energy of the X-rays exciting the sample was 160~\unit{\kilo\electronvolt}. We used the aforementioned 50$\times$ microscope objective to image the samples under X-ray excitation. For sample YAG:Yb PC-27 (see Sec.~\ref{sec:samples} for more information), this meant only 4--6 patterns out of 33 were in the field of view at any given time. Another complication of this experiment was that YAG:Yb does not scintillate as brightly as, e.g., YAG:Ce, while the quantum efficiency of the detector is $<10\%$ at 1030~\unit{\nano\meter}, the peak wavelength of YAG:Yb. These factors resulted in a lower than usual signal-to-noise ratio and very noticeable fluctuations in the intensity of scintillation. In fact, the signal was so low that we observed etaloning~\cite{etaloning} in our images. To address these limitations, we developed the following procedure for measuring scintillation enhancement:

\begin{enumerate}
    \item First, we moved the sample out of the field of view to measure the background. Using the software included with the micro-CT (Scout-and-Scan, version 16.2.18058.47373), we imaged the background and extracted a horizontal line profile. We repeated this three times and call the average line profile $\overline{B}(x)$, where $x$ is the coordinate along the line profile (or pixel number). The error in $\overline{B}(x)$ is taken to be the standard deviation of the three repetitions, $\Delta\overline{B}(x)$. To ensure consistency, all three line profiles were as long as possible (i.e., they spanned the full width of the image) and vertically centered with the image.
    \item We moved the sample into the field of view and, using a crosshair overlay in the software, vertically centered the patterns with the middle of the image. Then, we imaged the patterns and extracted a horizontal line profile three times, calling the average line profile $\overline{P}(x)$ (standard deviation $\Delta\overline{P}(x)$).
    \item The line profiles extracted in step 2 contain the intensity of light emitted by the patterns, but they are affected by etaloning. At the same time, scintillation enhancement is defined relative to the light emitted by the unpatterned regions on the surface of the sample. Therefore, to measure the scintillation enhancement while canceling out the effects of etaloning, we moved the patterns away from the vertical center of the image and, once again, imaged and extracted horizontal line profiles of the neighboring unpatterned region three times. It is important to note that this worked because the etaloning remains fixed and does not move with the sample. In other words, all three average line profiles $\overline{B}(x)$, $\overline{P}(x)$, and now $\overline{U}(x)$ experience the same modulation due to etaloning. This cancels out when calculating the spatially dependent scintillation enhancement:
    \begin{equation}
        \eta'(x)=\frac{\overline{P}(x)-\overline{B}(x)}{\overline{U}(x)-\overline{B}(x)}.
    \label{suppeq:eta_x}
    \end{equation}
    Data processing and propagation of uncertainty is further discussed in Sec. \ref{sec:data_processing}.
    \item We repeated steps 2--3 until we collected data for all 33 samples.
\end{enumerate}

All images used a 1~\unit{\second} exposure time and 4x4 binning. Typically, the widths of the line profiles (i.e., perpendicular to the direction in which they were drawn) were 16 px. This meant each line profile was actually the average of 16 parallel and adjacent line profiles, which effectively reduced noise. We occasionally set the line profile width to 24 or 32 px to span the full widths of the patterns, but this had no noticeable effect on the results.

\textit{Slit experiment:} The procedure for the micro-CT setup is identical to that for the UV laser setup. However, we collected fewer data points since using the micro-CT is much more time-consuming (for each data point, it involved turning the source on and off, opening the enclosure, manually moving the sample using the XY translation stage, etc.). We set the voltage and power of the source to 30 kVp and 2~\unit{\watt}, meaning $>99.9\%$ of the X-rays were blocked by the 1.2-\unit{\milli\meter}-thick steel slit (estimated using the Beer-Lambert law and approximating the mass attenuation coefficient of steel as that of iron). All the images used a 20~\unit{\second} exposure time.

It should be noted that the X-rays emerging from the source spread out, which resulted in geometric magnification. This was not strongly present in the the UV laser setup because the laser light was collimated. Geometric magnification is further discussed in Sec. \ref{sec:data_processing}, but the key takeaway is that the slit width was not necessarily equal to the width of the excited region of the scintillator.

\subsection{Synchrotron Setup}\label{subsec:synch}

\begin{figure}
    \centering
    \includegraphics[scale=1]{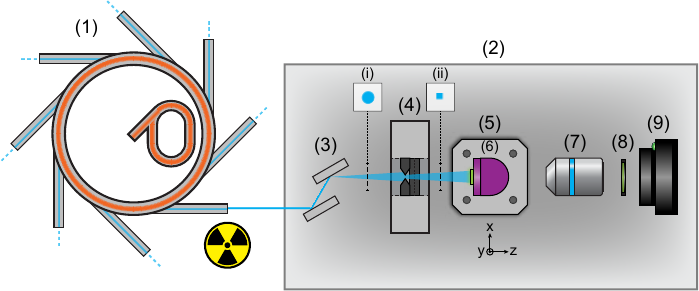}
    \caption{\textbf{Schematic of the synchrotron setup and its components.} (1) Synchrotron source. (2) Hutch. (3) Double crystal monochromator. (4) Adjustable slits. (5) 3-axis piezo stage. (6) 3D printed sample holder. (7) Objective. (8) Filter (9) Camera. (i) Beam cross-section before slit. (ii) Beam cross-section after slit.}
    \label{suppfig:synch}
\end{figure}

\begin{figure*}
    \centering
    \includegraphics{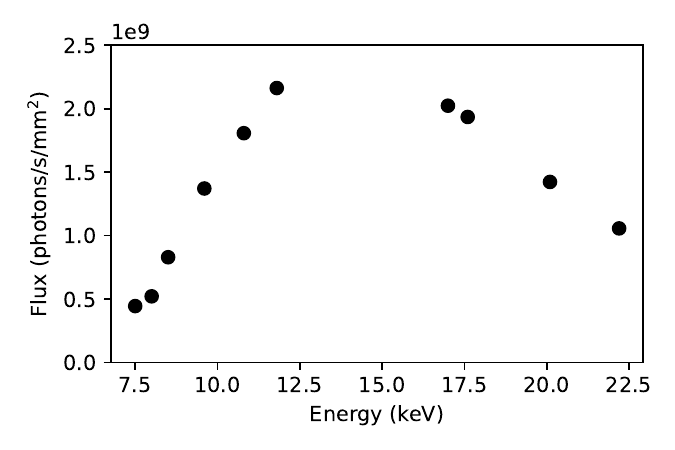}
    \caption{\textbf{Energy-dependent X-ray photon flux at the Diamond Light Source synchrotron.} Data were recorded using a calibrated diode in May 2025.}
    \label{suppfig:diamondflux}
\end{figure*}

Synchrotron measurements were carried out on the B16 beamline at the Diamond Light Source synchrotron over the course of two separate experiments: the YAG:Yb sample, PC-27, was characterized in October 2024, and the YAG:Ce sample, PC-26, in May 2025. Figure~\ref{suppfig:synch} shows the experimental setup during these measurements. The beam energy was selected using a double crystal monochromator; measurements were recorded at energies of either 8, 11, 14, 17, 20, 23, and 25~\unit{\kilo\eV} (YAG:Yb) or 8, 11.8, 17, 20.1 and 22.2~\unit{\kilo\eV} (YAG:Ce). In the May 2025 experiments, a calibrated diode was used to calculate the energy-dependent X-ray photon flux, shown in Fig.~\ref{suppfig:diamondflux}. The beam size in $x$ and $y$ (i.e., perpendicular to the beam itself) was controlled using two pairs of tungsten slits (JJ X-ray). For the YAG:Ce experiments, an additional set of slits was added further upstream to reduce diffraction from the slit edges. The scintillator was mounted on a stack of attocube piezo stages, enabling precise XYZ translation and rotation about three axes. An Optique Peter X-ray imaging microscope  was used to image the scintillator, consisting an objective lens with 10$\times$ magnification and a numerical aperture of 0.4. A bandpass filter was used to exclude emitted light of wavelengths outside the band of interest. For YAG:Ce, the filter had $>$~90\% transmission at the centre wavelength of 550~nm (Semrock FF01-550/200-25). For YAG:Yb, the filter had $>$~90\% transmission at the centre wavelength of 1030~nm (Thorlabs FLH1030-10). Images were captured using a Hamamatsu ORCA-Quest qCMOS camera, which has a conversion factor of 0.11 electrons/count and a typical quantum efficiency of 85\% at 550~nm and 3\% at 1030~nm. Lights in the hutch were switched off during measurements, and blackout materials were used to reduce stray light detected by the camera. 

\clearpage
\section{Imaging of the Volumetric Nanophotonic Scintillators}\label{sec:imaging}

In this section, we explain how we characterized the volumetric nanophotonic scintillators using multiple different imaging modalities.

\textit{Imaging under UV and X-ray excitation:} To obtain the images shown in Fig. 2(b) in the main text, which shows light emission from or around the individual holes that make up the patterns, we used the same experimental setups described in Secs. \ref{subsec:uv} and \ref{subsec:ct}. In the case of UV excitation, the sample was brought into focus by hand using the Z translation stage, while in the case of X-ray excitation, the computer-controlled sample stage was used. Both UV and X-ray excitation images were captured using the 50$\times$ objective.

\textit{Optical microscopy:} The optical microscope images shown in Fig. 1(c) in the main text were captured by standing the sample on its edge, using a piece of plastic to support it. Since the patterns were arranged in rows and the scintillator was transparent, we moved the microscope stage up and down to focus on a particular row of patterns. However, this made the tops of the patterns (i.e., the side on which they were etched) appear blurry because the edge of the sample, which is closer to the microscope objective, is out of focus. This created uncertainty in the depths of the patterns, which is further discussed in Sec. \ref{sec:data_processing}A.

\textit{Atomic force microscopy:} Atomic force microscope (AFM) images were captured using an Asylum Research Jupiter XR AFM in tapping mode. We used 160AC-NA tips from NanoAndMore USA. Figure 1(d) in the main text shows an AFM image of the cross-section of a typical pattern. This image was captured by embedding the broken sample in a piece of foam such that it stood on its edge, making the cross-section accessible to the tip. AFM images were postprocessed using Gwyddion version 2.66 (available at \url{https://gwyddion.net/})~\cite{Necas2012}. This included leveling the data and correcting horizontal scars.

\textit{Confocal microscopy:} We used a Keyence VK-X250 3D laser scanning confocal microscope to capture top (Fig. 1(b) in the main text) and cross-sectional (Fig. 1(d)(i)) views of the patterns. We embedded the broken sample in a piece of foam to view it from the side, same as the AFM images.

\clearpage
\section{Data Processing}\label{sec:data_processing}

\subsection{Estimation of Pattern Depth Using Optical Microscope Images}

\begin{figure}
    \centering
    \includegraphics{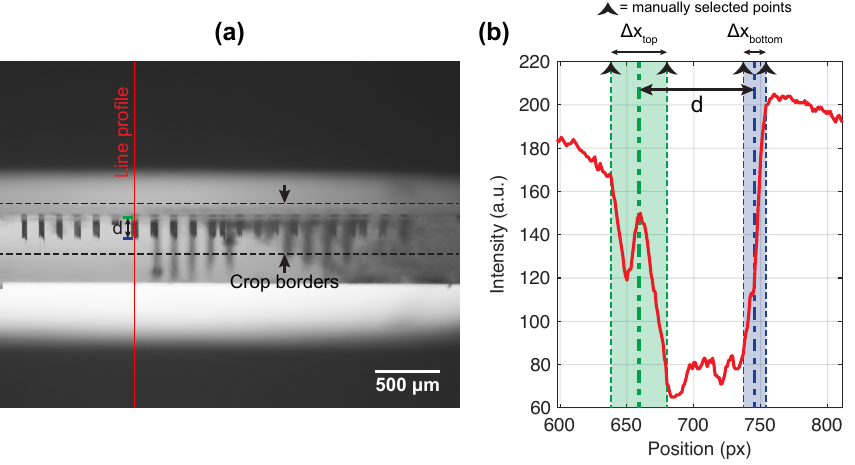}
    \caption{\textbf{Estimation of pattern depth.} In this example, we estimate the depth of pattern 16 in PC-27 (see Sec.~\ref{sec:samples}) after the third and final round of etching. (a) Optical microscope image, converted from RGB to grayscale. We draw a line profile through the center of the pattern (solid red line), then crop the line profile to zoom in on the pattern (crop borders indicated by black dashed lines; same for all patterns). The green and blue dashes indicate the top and bottom edges of the pattern, which we are about to determine. (b) Intensity along the line profile. By inspection, we manually select the edges of the blurry regions at the top (green) and bottom (blue) of the pattern. These points are indicated by the black wedges and used to calculate $\Delta x_{\text{top}}$, $\Delta x_{\text{bottom}}$, and $d$ as shown. The units can be converted from px to \unit{\micro\meter} using the scale bar. Finally, using Eq. (\ref{suppeq:depth_error}) to calculate $\Delta d$, we estimate the pattern depth to be $156.1\pm40.9$~\unit{\micro\meter}.}
    \label{suppfig:depth_estimation}
\end{figure}

To estimate the depths of the patterns, we used optical microscope images taken after each round of etching. We wrote a MATLAB script to load an optical microscope image and prompt the user to do the following:

\begin{enumerate}
    \item Draw a line profile through a pattern, as shown in Fig. \ref{suppfig:depth_estimation}(a). The line profile is plotted in another window in real time to help select the best data.
    \item Select the edges of the blurry regions located at top and bottom of the pattern, as shown in Fig. \ref{suppfig:depth_estimation}(b). Then, calculate the widths of the blurry regions $\Delta x_{\text{top}}$ and $\Delta x_{\text{bottom}}$, as well as the center-to-center distance between them, which is the estimated depth $d$.
    \item Estimate the uncertainty in $d$ by adding the blur widths in quadrature:
    \begin{equation}\label{suppeq:depth_error}
        \Delta d = \sqrt{\left(\Delta x_{\text{bottom}}\right)^2 + \left(\Delta x_{\text{top}}\right)^2}.
    \end{equation}
    \item Repeat steps 1--3 for all remaining patterns.
\end{enumerate}

Following this procedure, the estimated depths and their uncertainties, $d\pm\Delta d$, are plotted against scintillation enhancement in Fig. 3(c) in the main text. Histograms of $d$ after each round of etching are plotted in Fig. 1(c)(iii). These plots show that the depths of the patterns increase with time spent etching.

\subsection{Scintillation Enhancement}\label{subsec:enh_calc}

As previously mentioned, when measuring scintillation enhancement, we extracted three sets of line profiles to calculate the spatially dependent scintillation enhancement, $\eta'(x)$, given by Eq.~(\ref{suppeq:eta_x}). There were, however, additional steps involved.

\begin{figure}
    \centering
    \includegraphics{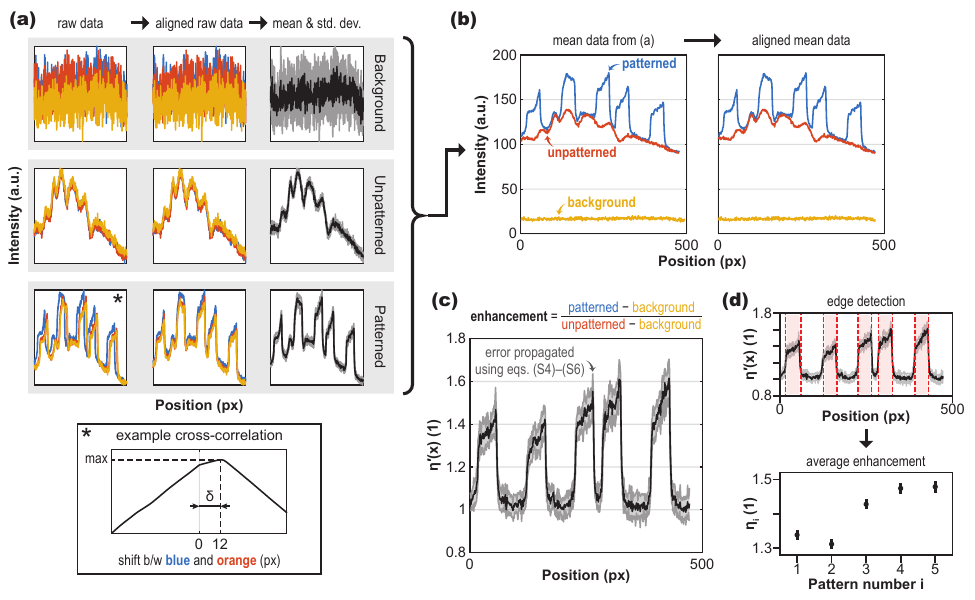}
    \caption{\textbf{Calculating scintillation enhancement.} (a) We align the three raw data curves using a cross-correlation approach. An example cross-correlation for the patterned signal (third row) is shown in the inset. Once aligned, we calculate the mean and standard deviation of each signal, shown by the shaded area. (b) The mean signals are then aligned relative to each other and cropped. In this example, the misalignment is relatively small, so the difference is not that noticeable. (c) We calculate the spatially-dependent emission enhancement and its error. (d) Using an edge-detection algorithm, we find the edges of each pattern and finally compute the spatially-averaged emission enhancement.}
    \label{suppfig:calc_enh}
\end{figure}

\textit{Alignment:} The line profiles were drawn by hand in Scout-and-Scan. As a result, they slightly varied in length and position. To align line profiles of the same kind (i.e., repeated measurements to be averaged), we used MATLAB to calculate the cross-correlation of each line profile relative to one chosen to be the reference. Then, we plotted the cross-correlation as a function of the lag. Since they represent roughly the same signal shifted in space, we expect the cross-correlation of two line profiles to have a single peak. The value of the lag at this peak is the amount by which we need to shift one of the line profiles to align it with the other. After aligning the line profiles, they were cropped to have the same length. This fully automated approach produced accurate results and, compared to aligning the line profiles by hand, was fast and deterministic. This step is shown in Fig.~\ref{suppfig:calc_enh}(a).

To align different kinds of line profiles (i.e., patterned, unpatterned, and background), we used the same approach and took advantage of the fact that etaloning was observed in all three signals. In most cases, calculating the cross-correlations of the full line profiles worked even though the patterned signal jumps where patterns are located. This was likely because the patterns are small compared to the total length of the line profile. When using the full line profiles did not work, we either cropped the patterned signal to a small region between patterns with noticeable etaloning and tried again or aligned the line profiles by hand (which we rarely had to do). Again, the line profiles were cropped after alignment. This step is shown in Fig.~\ref{suppfig:calc_enh}(b).

\textit{Calculating scintillation enhancement:} Eq.~(\ref{suppeq:eta_x}) is the spatially dependent scintillation enhancement, calculated using the average background, patterned, and unpatterned line profiles. This is shown in Fig.~\ref{suppfig:calc_enh}(c). By definition, $\eta'(x)\approx1$ everywhere except where patterns are located, so the average scintillation enhancement of pattern $i$ ($i=1,2,3,\ldots$) is

\begin{equation}\label{suppeq:etabar}
    \eta_{i} = \frac{1}{x_{i,R} - x_{i,L}}\int_{x_{i,L}}^{x_{i,R}}\eta'(x)dx\approx\frac{1}{N}\sum_{x_{j}=x_{i,L}}^{x_{i,R}}\eta'(x_{j}),
\end{equation}

where $x_{i,L}$ and $x_{i,R}$ are the locations of the left and right edges of the pattern, respectively, $N=x_{i,R}-x_{i,L}$, and $j$ denotes pixel number (since in reality, the data are discrete, not continuous). The edges were automatically detected using \verb|ruptures| (version 1.1.9), a Python library for change point detection~\cite{ruptures}. This step is shown in Fig.~\ref{suppfig:calc_enh}(d), followed by the pattern-averaged values $\eta_i$. Combined with our MATLAB script for aligning and averaging the line profiles, this resulted in a fully automated data processing pipeline.

\textit{Propagation of uncertainty:} As previously mentioned, the average line profiles $\overline{B}(x)$, $\overline{P}(x)$, and $\overline{U}(x)$ each had their own standard deviations. The error in $\eta'(x)$ was calculated in two steps:

\begin{enumerate}
    \item Calculate the relative errors in the numerator and denominator of Eq.~(\ref{suppeq:eta_x}) by adding the standard deviations in quadrature:
    \begin{equation}\label{suppeq:num_error}
        \delta\,\text{num}\left[\eta'(x)\right]=\frac{\sqrt{\Delta\overline{P}(x)^2+\Delta\overline{B}(x)^2}}{\overline{P}(x)-\overline{B}(x)},
    \end{equation}
    \begin{equation}\label{suppeq:den_error}
        \delta\,\text{den}\left[\eta'(x)\right]=\frac{\sqrt{\Delta\overline{U}(x)^2+\Delta\overline{B}(x)^2}}{\overline{U}(x)-\overline{B}(x)},
    \end{equation}
    where $\text{num}(\frac{a}{b})=a$ and $\text{den}(\frac{a}{b})=b$ denote the operation of taking the numerator and denominator, respectively. 
    \item Calculate the absolute error in $\eta'(x)$ by adding these relative errors in quadrature,
    \begin{equation}\label{suppeq:eta_x_error}
        \Delta \eta'(x)=\eta'(x)\sqrt{\delta\,\text{num}\left[\eta'(x)\right]^2 + \delta\,\text{den}\left[\eta'(x)\right]^2}.
    \end{equation}
\end{enumerate}

To calculate the error in $\eta_i$, we added two components in quadrature. First, we calculated the standard error of the mean to account for spatial variations in $\eta'(x)$ (e.g., due to the non-uniformity of the patterns):

\begin{equation}\label{suppeq:mean_error}
    \Delta_{\text{mean}}(\eta_{i})=\frac{\sigma_{i}}{\sqrt{N}}=\sqrt{\frac{1}{N}\sum_{x_{j}=x_{i,L}}^{x_{i,R}}\left[\eta'(x_{j})-\eta_i\right]^2},
\end{equation}

where $\sigma_i$ is the standard deviation of $\eta'(x)$ evaluated from $x_{i,L}$ to $x_{i,R}$. This is the error associated with averaging $\eta'(x)$ itself. The error associated with repeated measurements, $\Delta \eta'(x)$, was added in quadrature and divided by $N$:

\begin{equation}\label{suppeq:repeat_error}
    \Delta_{\text{meas.}}(\eta_i) = \frac{1}{N}\sqrt{\sum_{x_{j}=x_{i,L}}^{x_{i,R}}\Delta\eta'(x_{j})^2}.
\end{equation}

This can be understood by inspecting Eq. (\ref{suppeq:etabar}): $\Delta \eta(x_j)$ is added in quadrature due to the sum over $j$ when averaging, then divided by the scalar $1/N$.  Finally, we calculated the error in $\eta_i$ by adding these in quadrature:

\begin{equation}\label{suppeq:etabar_error}
    \Delta\eta_i = \sqrt{\Delta_{\text{mean}}(\eta_i)^2 + \Delta_{\text{meas.}}(\eta_i)^2}.
\end{equation}

Eq.~(\ref{suppeq:etabar_error}) accounts for both spatial variations in scintillation enhancement on each pattern ($\Delta_{\text{mean}}(\eta_i)$) and temporal variations due to signal fluctuations ($\Delta_{\text{meas.}}(\eta_i)$).

\subsection{Remote Outcoupling}\label{subsec:lc_calc}

In the slit experiments, our goal was to observe and attempt to quantify the patterns' ability to remotely outcouple light emitted far away in the bulk, without being directly excited themselves. From the images captured as a function of slit width $w=2a$ and slit-to-pattern distance $y_i$ ($i=1,2,3,\ldots$ denotes the pattern number), for each pattern, we calculated the average intensity of remotely outcoupled light,
\begin{equation}\label{suppeq:lc}
    \overline{I}_i = \frac{1}{A_{i}}\iint_{A_i}\left[I_{i}(x,y) - I_{\text{dark}}(x,y)\right]dA,
\end{equation}
where $A_i$ is the area of pattern $i$, $I_{i}(x,y)$ is the spatially dependent intensity, and $I_{\text{dark}}(x,y)$ is the dark frame. This is relatively straightforward, but a challenge was that in some experiments---specifically, those involving the UV laser setup described in \ref{subsec:uv}---the slit was fixed in place and the patterns were moved relative to it to control $y_i$. This meant that, from image to image, the absolute positions of the patterns changed. To deal with this, we developed the following procedure in MATLAB:
\begin{enumerate}
    \item Load an image in which the position of the slit is halfway between two rows of patterns (i.e., the initial position described in \ref{subsec:uv}). Let $I_{\text{ref}}(x,y)$ be the intensity at coordinates $(x,y)$. Using this image as a reference, draw a rectangular region of interest (ROI) around each pattern. The ROI should not include any pixels outside the pattern. The ROI can be thought of as a binary mask or 2D boxcar function that crops the image:
    \begin{equation}
    \text{ROI}_{i}(x,y) = \text{rect}\left[\frac{x-\left(\frac{x_{i,R}+x_{i,L}}{2}\right)}{x_{i,R}-x_{i,L}}\right]\text{rect}\left[\frac{y-\left(\frac{y_{i,T}+y_{i,B}}{2}\right)}{y_{i,T}-y_{i,B}}\right],
    \end{equation}
    where
    \begin{equation}
        \text{rect}(x)=
        \begin{cases}
            1 & \text{if } |x|<\frac{1}{2}, \\
            0 & \text{if } |x|>\frac{1}{2}.
        \end{cases}
    \end{equation}
    $x_{i,L}$ and $x_{i,R}$ are the left and right edges of pattern $i$, respectively; $y_{i,T}$ and $y_{i,B}$ are the top and bottom edges. These ``pattern ROIs,'' including the pattern intensities from the reference image,
    \begin{equation}
        I_{i,\text{ref}}(x,y)=I_{\text{ref}}(x,y)\text{ROI}_{i}(x,y),
    \end{equation}
    are stored in a \verb|.mat| file. This is done only once per set of experiments.
    \item In another script, load the dark frame $I_{\text{dark}}(x,y)$ and the pattern ROIs, which are pre-defined as per step 1. The images are stored in separate folders depending on the parameter being swept ($w$ or $y_i$). Select a folder.
    \item Now, sweeping over each image in the folder:
    \begin{enumerate}
        \item For each $w$-value, draw the ``slit ROI,'' which is done using a fully automated approach. The left and right edges of the slit ROI, $x_{\text{slit},L}$ and $x_{\text{slit},R}$, are the same for all images. Draw a vertical line profile that passes through the slit; then, use \verb|findchangepts| to automatically detect the top and bottom edges of the slit, $y_{\text{slit},T}$ and $y_{\text{slit},B}$. This is done only once for datasets where $w$ is fixed and $y_i$ is free. For datasets where $w$ is free and $y_i$ is fixed, this is done for each image in the folder.
        
        \textbf{Note:} in the case of the synchrotron experiments on YAG:Yb PC-27, this approach was not used. There were so few $w$-values that it was more efficient to draw the slit ROIs by hand only once and reuse them, as was done for the pattern ROIs (see step 1).
        \item If needed (i.e., for datasets where $y_i$ is free and $w$ is fixed), align the pattern ROIs with the patterns in the current image $I(x,y)$ using a fully automated cross-correlation approach similar to the one described in \ref{subsec:enh_calc}. In this case, for each pattern $i$, the image is cropped to the width of the pattern ROI:
        \begin{equation}
            I_{i,\text{crop}}=I(x,y)\text{rect}\left[\frac{x-\left(\frac{x_{i,R}+x_{i,L}}{2}\right)}{x_{i,R}-x_{i,L}}\right].
        \end{equation}
        Then, each row of $I_{i,\text{ref}}(x,y)$, which is denoted by $j$, is cross-correlated with $I_{i,\text{crop}}$. This results in a lag value for each row corresponding to the maximum value of the cross-correlation, $\tau_j$. To align the pattern ROI with the patterns in the image, it is shifted along the $y$-coordinate by the mode of the $\tau_j$-values, $\hat{\tau}$. This allows us to calculate the pattern intensities:
        \begin{equation}
            I_{i}(x,y)=I(x,y)\text{ROI}_{i}(x,y-\hat{\tau}).
        \end{equation}
        \textbf{Note:} in the case of the micro-CT experiments on YAG:Yb PC-27, this approach did not work because the signal-to-noise ratio was too low. Instead, the pattern ROIs were carefully aligned by hand. Since all the patterns on this sample are in a row, an arbitrary pattern ROI was aligned with the corresponding pattern in the image, and the shift along the $y$-coordinate is assumed to be identical for the remaining pattern ROIs. In the case of the synchrotron experiments on YAG:Ce PC-24, the patterns remained still in the field of view and the slit moved, meaning there was no need to correct the positions of the patterns, and $y_i$ could be determined using an edge detection algorithm on the slit with the fixed pattern positions.
        \item Calculate $y_i$:
        \begin{equation}
            y_i = \left|\left(y_{i,B}+\frac{y_{i,T}-y_{i,B}}{2}+\hat{\tau}\right) - \left(y_{\text{slit},B}+\frac{y_{\text{slit},T}-y_{\text{slit},B}}{2}\right)\right|.
        \end{equation}
        This, along with $w$, are converted to SI units using a scale bar. The scale bar was calculated using an image of a USAF test target in the case of the UV laser setup and the known side lengths of the patterns in the case of the micro-CT setup. It is important to note that $w$, which we call the slit width, is actually the width of the excited region of the scintillator. $w$ is directly proportional to the true slit width, $w_{\text{true}}$. In the synchrotron setup and the UV laser setup, $w\approx w_{\text{true}}$ because the beam was either narrow or collimated. In the micro-CT setup, the X-rays are divergent, resulting in geometric magnification such that $w=M_{g}w_{\text{true}}$, where $M_{g}=x_{s}/x_{o}$, $x_s$ is the source-to-scintillator distance, and $x_o$ is the source-to-slit distance~\cite{MartinMonier2024}. $M_g$ can be directly calculated using measurements of $x_o$ and $x_s$ or inferred from $w=y_{\text{slit},T}-y_{\text{slit},B}$ and the micrometer reading recorded in the filename of each image ($w_{\text{true}}$) since $x_{s}/x_{o}=w/w_{\text{true}}$ using similar triangles. Using a photograph of the micro-CT setup to estimate $x_o$ and $x_s$, we found that $M_g\approx4.2$. This was corroborated by a calculation of $w/w_{\text{true}}$ using an arbitrary image.
        \item Calculate $\overline{I}_i$ using Eq. (\ref{suppeq:lc}) and the standard deviations of $I_{i}(x,y)$ and $I_{\text{dark}}(x,y)$ ($\Delta I_i$ and $\Delta I_{\text{dark}}$, respectively). Then the absolute error in $\overline{I}_i$ is
        \begin{equation}
            \Delta\overline{I}_i = \sqrt{(\Delta I_i)^2 + (\Delta I_{\text{dark}})^2}.
        \end{equation}
        \item When a sample has more than one pattern on it (e.g., in the case of PC-27), we calculate the best estimate of the mean pattern intensity:
        \begin{equation}\label{eq:est_mean}
            \overline{I}_{p} = \frac{\sum_{i=1}^{k}\frac{\overline{I}_i}{(\Delta\overline{I}_i\,)^2}}{\sum_{i=1}^{k}\frac{1}{(\Delta\overline{I}_i\,)^2}},
        \end{equation}
        where $k$ is the number of patterns. The best estimate of the standard deviation of the mean pattern intensity is:
        \begin{equation}\label{eq:est_stdev}
            \Delta \overline{I}_p = \sqrt{\frac{1}{\sum_{i=1}^{k}\frac{1}{(\Delta\overline{I}_i\,)^2}}}.
        \end{equation}
        Obviously, if $k=1$ (only one pattern on the sample), $\overline{I}_p =\overline{I}_1$ and $\Delta\overline{I}_p = \Delta\overline{I}_1$~\cite{Robinson2016}. Additionally, we calculate the average slit-to-pattern distance $y_p$ since the patterns are almost always in a row and variations in $y_i$ are relatively small. Altogether, calculating $I_p$ and $\Delta I_p$  consolidates multiple different measurements of light collecting into a single data point $(w,y_p)$.
    \end{enumerate}
    \item Finally, save the results ($w$, $y_p$, $\overline{I}_p$, and $\Delta\overline{I}_p$, as well as $E$ in the case of the synchrotron experiments).
\end{enumerate}

\textit{Data fitting:} After data processing, we used the \verb|scipy.optimize| package in Python to fit $\overline{I}_i$ to Eq. (3) in the main text using the least squares method (implemented via the \verb|least_squares| function). In the case of the synchrotron experiments, where $E$ was also an independent variable in addition to $w$ and $y_i$, separate $\lambda$- and $C$-values were fitted for each $E$-value.

In Fig. 4(c)(i) in the main text, we show the values of the fit parameters $\lambda$ and $C$, with the error bars representing their respective standard errors. These standard errors were calculated by taking the square roots of the diagonal components of the covariance matrix:
\begin{equation}
    \boldsymbol{\Sigma} = \hat{\sigma}^2(\mathbf{J}^T\mathbf{J})^{-1},
\end{equation}
where $\mathbf{J}$ is the Jacobian matrix, and $\hat{\sigma}^2$ is the residual variance, estimated as the sum of squared residuals divided by the degrees of freedom $\nu=N-p$, with $N$ data points and $p=2$ fitting parameters. These quantities ($\mathbf{J}$, the sum of squared residuals, and $\nu$) are either outputs of \verb|least_squares| or are readily calculated from its outputs.

We also show the uncertainty in the curve fits as a shaded area representing the pointwise 95\% confidence interval. If $\hat{I}_E(a;y_0)$ is the curve fitted to the experimental data (i.e., the predicted value), the upper and lower limits of the 95\% confidence interval are given by
\begin{equation}\label{eq:CI}
    CI = \hat{I}_E(a;y_0) \pm t_{(1+CL)/2,\,\nu}\cdot\hat{\sigma}_f,
\end{equation}
where $t_{(1-CL)/2,\,\nu}$ is Student's t-distribution evaluated at a confidence level $CL=0.95$ and degrees of freedom $\nu$, and $\hat{\sigma}_f$ is the estimated standard error of the fitted curve $\hat{I}_E(a;y_0)$~\cite{NISThandbook,SeberWild}. We calculate $\hat{\sigma}_f$ using the following expression:
\begin{equation}
    \hat{\sigma}_f = \sqrt{\textbf{g}^{T}\boldsymbol{\Sigma}\textbf{g}},
\end{equation}
where
\begin{equation}
    \textbf{g} = 
    \begin{bmatrix}
        \partial \hat{I}_E/\partial \lambda \\
        \partial \hat{I}_E/\partial C
    \end{bmatrix}
\end{equation}
is the gradient of $\hat{I}_E(a;y_0)$ with respect to the fit parameters~\cite{SeberWild}. We computed the derivative with respect to $\lambda$ using a central finite difference, while the derivative with respect to $C$ can be found analytically: $\partial \hat{I}_E/\partial C = \hat{I}_E(a;y_0)/C$. The 95\% confidence intervals given by Eq. (\ref{eq:CI}) and plotted in Fig. 4(c) should be seen as the estimated, pointwise uncertainties for the fitted curves, not as prediction intervals for the experimental data.

Finally, although the mean and standard deviations calculated according to Eqs. (\ref{eq:est_mean}) and (\ref{eq:est_stdev}) are plotted in Fig. 4(c) in the main text to consolidate a large amount of data, the data are fitted treating each pattern and pair of $(w, y_i)$ as an individual data point.

% These were calculated by taking the square roots of the diagonal components of the covariance matrix, which are the standard errors in the parameter estimates, $\sigma_{\lambda}$ and $\sigma_{C}$, then using the formula
% \begin{equation}
%     \begin{bmatrix}
%         \lambda_\pm \\
%         C_\pm
%     \end{bmatrix} =
%     \begin{bmatrix}
%         \lambda \\
%         C
%     \end{bmatrix} \pm
%     z \begin{bmatrix}
%         \sigma_\lambda \\
%         \sigma_C
%     \end{bmatrix},
% \end{equation}
% where $z=\text{PPF}(1-\frac{1-\text{CL}}{2})$ is the percent point function (hence, PPF) of the normal distribution, and $\text{CL}=95\%$ is the confidence level. The $+$ and $-$ subscripts correspond to the upper and lower confidence intervals, respectively.

\clearpage
\section{Samples Studied in This Work}\label{sec:samples}

The results shown in the main text are related to multiple different samples, which we describe in this section.

\begin{figure}
    \centering
    \includegraphics[scale=2]{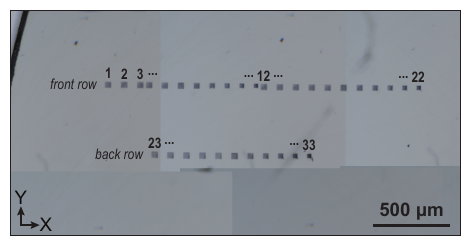}
    \caption{\textbf{Top view optical microscope image of PC-27 (cf. Table \ref{tab:pc27}).} This is a composite image consisting of 6 separate images stitched together. This sample was used to measure remote outcoupling in Fig.~4(c), bottom panels. Additional images of PC-27 are shown in Fig.~1(c)(i--ii). Patterns 23--27 can be seen in Fig.~3(a).}
    \label{suppfig:pc27}
\end{figure}

One of the most important samples was ``PC-27,'' which was made of YAG:Yb (peak emission wavelength 1030~\unit{\nano\meter}). The bulk scintillator was sourced from Optogama. This sample consisted of 33 patterns with different periods and nominal depths written by the laser. As discussed in the main text, the actual depths of the patterns depended on the amount of time spent etching (which we divided into three etching sessions lasting 5 h each, with a round of imaging in between). In addition to period and nominal depth, the patterns were characterized by the power of the laser during writing. PC-27 was an important sample because it swept over several key parameters related to scintillation enhancement, as shown in Fig.~3 in the main text. Table \ref{tab:pc27} summarizes the properties (period, nominal depth, and writing power) of the patterns on PC-27. For comparison, a top view optical microscope image is shown in Fig. \ref{suppfig:pc27}. The ``front row'' of patterns (1--22) can be seen in Fig.~1(c)(i) in the main text, while the ``back row'' (23--33) can be seen in Fig.~1(c)(ii). Additionally, patterns 23--27 are shown in the example X-ray excitation image in Fig. 3(a).

Another sample used in the synchrotron slit experiments was ``PC-24,'' made of YAG:Ce (peak emission wavelength 550~\unit{\nano\meter}) sourced from Crytur. This sample had 16 patterns on it, but only 4 were used. The properties of these patterns are summarized in Table \ref{tab:pc24}.

In the main text, the sample seen in Figs. 1(b) and 2(b) is ``PC-16,'' made of YAG:Ce, also sourced from Crytur. This sample had 7 patterns on it, but only one (pattern 2) was considered to be of high enough quality to be included in the results of the slit experiments. Some of the patterns are ``disordered,'' meaning the $x$ and $y$ coordinates of each lattice site are randomly perturbed by some displacement $\delta$. The properties of these patterns are summarized in Table \ref{tab:pc16}.

Finally, the sample seen in Fig. 1(d) is ``PC-14,'' made of YAG:Ce as well. This sample broke into two pieces by accident, but it was still used to image the cross-section of one of the patterns. The properties of these patterns are summarized in Table \ref{tab:pc14}.

\clearpage
% ========================
% ========================
% ~~~  PC-27  ~~~
% ========================
% ========================
\begin{longtable}{|c|c|c|c|>{\centering\arraybackslash}m{2in}|}
\caption{Patterns on YAG:Yb PC-27 (cf. Fig. \ref{suppfig:pc27})}\label{tab:pc27} \\
% == First heading. ===
\hline
Pattern no. & Period (\unit{\micro\meter}) & Nominal depth (\unit{\micro\meter}) & Power before lens (\unit{\milli\watt}) & Relevant main text figures \\
\hline
% \hline
\endfirsthead
% =======================

% === Second heading. ===
\multicolumn{5}{c}
{{TABLE \thetable\ (continued): Patterns on YAG:Yb PC-27}} \\
\hline
Pattern no. & Period (\unit{\micro\meter}) & Nominal depth (\unit{\micro\meter}) & Power before lens (\unit{\milli\watt}) & Relevant main text figures \\
\hline
% \hline
\endhead
% =======================

% === Footers for split table. ===
\hline
\multicolumn{5}{r}{{Continued on next page}} \\
\endfoot

\hline
\endlastfoot
% =======================

% ~~~ First table page. ~~~
\multicolumn{5}{|c|}{{\textbf{Front row of patterns}}} \\
\hline
1 & \multirow{6}{*}{1.5} & 600 & \multirow{11}{*}{90/200} & \multirow{22}{=}{\centering 1(c)(i), 4(c) bottom panels.} \\ 
2 &  & 500 & & \\  
3 &  & 400 & & \\
4 &  & 300 & & \\
5 &  & 200 & & \\
6 &  & 100 & & \\
\cline{1-3}
7 & 1.4 & \multirow{5}{*}{600} & & \\
8 & 1.3 & & & \\
9 & 1.2 & & & \\
10 & 1.1 & & & \\
11 & 1.0 & & & \\
\cline{1-4}
12 & \multirow{6}{*}{1.5} & 600 & \multirow{11}{*}{100/300} & \\ 
13 &  & 500 & & \\  
14 &  & 400 & & \\
15 &  & 300 & & \\
16 &  & 200 & & \\
17 &  & 100 & & \\
\cline{1-3}
18 & 1.4 & \multirow{5}{*}{600} & & \\
19 & 1.3 & & & \\
20 & 1.2 & & & \\
21 & 1.1 & & & \\
22 & 1.0 & & & \\
\hline
\pagebreak
% ~~~~~~~~~~~~~~~~~~~~~~~

% ~~~ Second table page. ~~~
\multicolumn{5}{|c|}{{\textbf{Back row of patterns}}} \\
\hline
23 & \multirow{6}{*}{1.5} & 600 & \multirow{11}{*}{95/250} & \multirow{5}{=}{\centering 1(c)(ii), 3(a)--(c), 4(c) bottom panels.} \\ 
24 &  & 500 & & \\  
25 &  & 400 & & \\
26 &  & 300 & & \\
27 &  & 200 & & \\
\cline{5-5}
28 &  & 100 & & \multirow{6}{=}{\centering  1(c)(ii), 3(b) and (c), 4(c) bottom panels.} \\
\cline{1-3}
29 & 1.4 & \multirow{5}{*}{600} & & \\
30 & 1.3 & & & \\
31 & 1.2 & & & \\
32 & 1.1 & & & \\
33 & 1.0 & & & \\
\hline
% ~~~~~~~~~~~~~~~~~~~~~~~
\end{longtable}

% ========================
% ========================
% ~~~  PC-24  ~~~
% ========================
% ========================
\begin{longtable}{|c|c|c|c|>{\centering\arraybackslash}m{2in}|}
\caption{Patterns on YAG:Ce PC-24 (cf. Fig. \ref{suppfig:pc24})}\label{tab:pc24} \\
% == First heading. ===
\hline
Pattern no. & Period (\unit{\micro\meter}) & Nominal depth (\unit{\micro\meter}) & Power before lens (\unit{\milli\watt}) & Relevant main text figures \\
\hline
% \hline
\endfirsthead
% =======================

% == Second heading. ===
\multicolumn{5}{c}
{{TABLE \thetable\ (continued): Patterns on YAG:Ce PC-24}} \\
\hline
Pattern no. & Period (\unit{\micro\meter}) & Nominal depth (\unit{\micro\meter}) & Power before lens (\unit{\milli\watt}) & Relevant main text figures \\
\hline
% \hline
\endhead
% =======================

% === Footers for split table. ===
\hline
\multicolumn{5}{r}{{Continued on next page}} \\
\endfoot

\hline
\endlastfoot
% =======================

13 & \multirow{4}{*}{1.0} & 100 & \multirow{4}{*}{85/85} & \multirow{4}{=}{\centering  4(c) top-center panel.} \\ 
14 &  & 150 & & \\  
15 &  & 200 & & \\
16 &  & 250 & & \\
\hline

\end{longtable}

\begin{figure}
    \centering
    \includegraphics[scale=2]{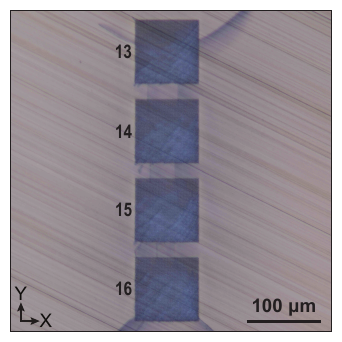}
    \caption{\textbf{Top view optical microscope image of PC-24 (cf. Table \ref{tab:pc24})} This sample was used to measure remote outcoupling in Fig.~4(c), top-center panel.}
    \label{suppfig:pc24}
\end{figure}

\clearpage
% ========================
% ========================
% ~~~  PC-16  ~~~
% ========================
% ========================
\begin{longtable}{|c|c|c|>{\centering\arraybackslash}m{2in}|}
\caption{Patterns on YAG:Ce PC-16 (cf. Fig. \ref{suppfig:pc16})}\label{tab:pc16} \\
% == First heading. ===
\hline
Pattern no. & Period (\unit{\micro\meter}) & $\delta$ (\unit{\micro\meter}) & Relevant main text figures \\
\hline
% \hline
\endfirsthead
% =======================

% == Second heading. ===
\multicolumn{4}{c}
{{TABLE \thetable\ (continued): Patterns on YAG:Ce PC-16}} \\
\hline
Pattern no. & Period (\unit{\micro\meter}) & $\delta$ (\unit{\micro\meter}) & Relevant main text figures \\
\hline
% \hline
\endhead
% =======================

% === Footers for split table. ===
\hline
\multicolumn{4}{r}{{Continued on next page}} \\
\endfoot

\hline
\endlastfoot
% =======================

1 & \multirow{5}{*}{1.0} & 0.25 & \makecell[c]{4(b)(ii) and (c),\\top-left and top-right panels.} \\
\cline{4-4}
2 & & $1/6$ & \multirow{3}{=}{\centering 4(b)(ii)} \\
3 & & 0.40 & \\
4 & & 0.40 & \\
\cline{4-4}
5 & & & \\
\cline{1-2} \cline{4-4}
6 & 1.5 & & 1(b), 2(b) \\
\cline{4-4}
7 & 2.0 & & \\
\hline

\end{longtable}

\begin{figure}
    \centering
    \includegraphics[scale=2]{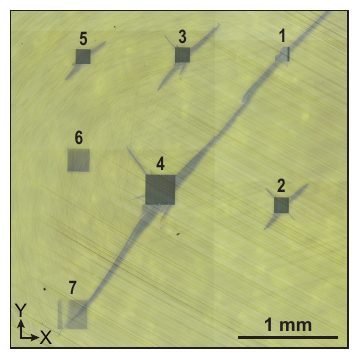}
    \caption{\textbf{Top view optical microscope image of PC-16.} This is a composite image consisting of 8 separate images stitched together. This sample, specifically pattern 2, was used to measure remote outcoupling in Fig. 4(c), top-left and top-right panels. Patterns 1--4 can be seen in Fig. 4(b)(ii), rotated counterclockwise by $90^{\circ}$. Additional images of pattern 6 are shown in Figs. 1(b) and 2(b).}
    \label{suppfig:pc16}
\end{figure}

% ========================
% ========================
% ~~~  PC-14  ~~~
% ========================
% ========================
\begin{longtable}{|c|c|c|}
\caption{Patterns on YAG:Ce PC-14 (cf. Fig. \ref{suppfig:pc14})}\label{tab:pc14} \\
% == First heading. ===
\hline
Pattern no. & Period (\unit{\micro\meter}) & Relevant main text figures \\
\hline
% \hline
\endfirsthead
% =======================

% == Second heading. ===
\multicolumn{3}{c}
{{TABLE \thetable\ (continued): Patterns on YAG:Ce PC-14}} \\
\hline
Pattern no. & Period (\unit{\micro\meter}) & Relevant main text figures \\
\hline
\hline
\endhead
% =======================

% === Footers for split table. ===
\hline
\multicolumn{3}{r}{{Continued on next page}} \\
\endfoot

\hline
\endlastfoot
% =======================

1 & \multirow{2}{*}{1.0} & 1(d) \\
\cline{3-3}
2 & &  \\
\cline{1-2}
3 & 1.2 & \\
4 & 1.4 & \\
\hline

\end{longtable}

\begin{figure}
    \centering
    \includegraphics{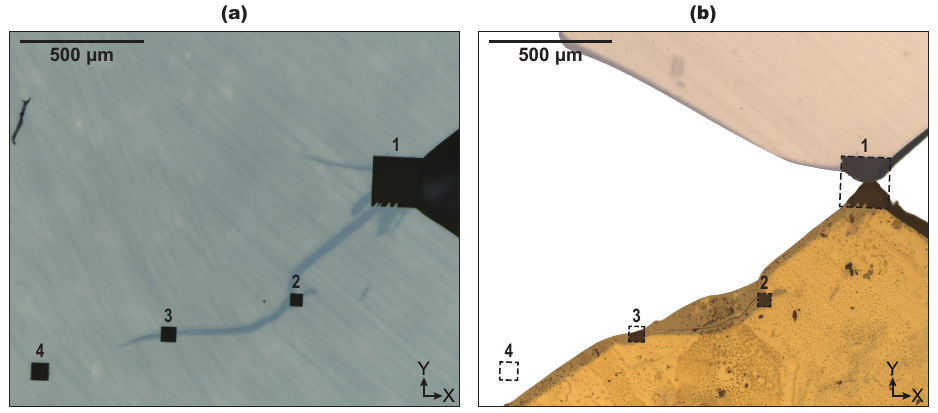}
    \caption{\textbf{Top view optical microscope image of PC-14.} (a) Before the sample fractured. (b) After the sample fractured, reconstructed from 2 separate images. Images of the cross-section of pattern 1 are shown in Fig. 1(d).}
    \label{suppfig:pc14}
\end{figure}

\clearpage
\section{Simple Model for Remote Outcoupling}\label{sec:model}

\begin{figure}[h]
    \centering
    \includegraphics{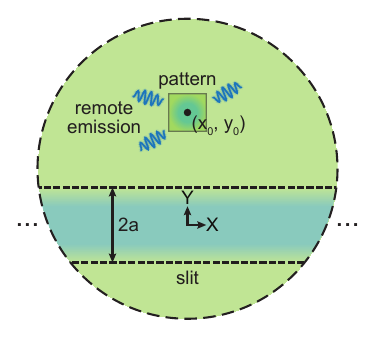}
    \caption{\textbf{Model for remote light collection in volumetric nanophotonic scintillators.} The scintillator is irradiated by high energy excitation through a slit of width $2a$. The PhC pattern is assumed to act as a point detector located at $(x_0, y_0)$.}
    \label{suppfig:modelfig}
\end{figure}

In this section, we develop a simple model for explaining and understanding the remote light collection ability of volumetric nanophotonic scintillators through a slit excitation, as observed in our experiments. X-ray or UV radiation generates a density of isotropic emitters in the slit region whose emission propagates through the scintillator while undergoing absorption and scattering. In the absence of nanophotonic structuring, a significant fraction of this laterally propagating light remains trapped by total internal reflection at the material-air interface. The embedded volumetric photonic crystal (PhC) pattern, however, provides additional scattering channels that enable otherwise confined or absorbed photons to couple out of the material. In this way, the PhC functions as a localized outcoupling site allowing for remote detection.

To model this behavior, we make the following assumptions:
\begin{enumerate}
    \item The slit is effectively infinite in extent in one direction (here along $x$).
    \item The high energy excitation through the slit produces a uniform density of  isotropic dipole emitters in the same region of the scintillator.
    \item The photonic crystal (PhC) pattern is a point object.
    \item The scintillator with the embedded PhC extends uniformly to infinity in the out-of-plane ($z$) direction, which reduces the problem to a two dimensional (2D) calculation.
\end{enumerate}

Under these assumptions, we define parameters as shown in Fig.~\ref{suppfig:modelfig}: the slit extends from $x \in (-\infty, \infty), y \in (-a, a)$ and the PhC is a point object located at coordinates $(x_0, y_0)$. 

The high energy excitation with energy $E$ creates a uniform areal density of isotropic 2D emitters $\sigma(E)$. For such a source, the radiated power spreads cylindrically, such that the intensity at distance $r$ scales as $\sim 1/r$. Assuming an exponential attenuation length of emission ($\lambda$), the 2D propagation kernel (Green's function) is
\begin{equation}
    g(r) = \frac{e^{-r/\lambda}}{2\pi r}.
\end{equation}
Here, $\lambda$ has the physical meaning of the mean-free path of the emitted photons, determined by a combination of material absorption and scattering.

If each emitter has emitted power $P_0$ into the 2D plane, then a differential area element, $dA$, contributes the following intensity at point $(x_0, y_0)$:
\begin{equation}
    dI_E = P_0\sigma(E) g(r)dA = \frac{P_0\sigma (E)}{2\pi} \frac{e^{-r/\lambda}}{r}dxdy,
\end{equation}
where $r = \sqrt{(x-x_0)^2+(y-y_0)^2}$.

Therefore, the total intensity collected at $(x_0, y_0)$, assuming a collection efficiency of $\varepsilon \in [0, 1]$ is given by
\begin{equation}
    I_E(a; y_0) = \frac{\varepsilon P_0\sigma (E)}{2\pi} \int_{-a}^a \int_{-\infty}^{\infty} \frac{e^{-\sqrt{(x-x_0)^2+(y-y_0)^2}/\lambda}}{\sqrt{(x-x_0)^2+(y-y_0)^2}}dx dy.
\end{equation}

Solving the integral in $x$ using the expression $\int_0^{\infty}e^{-\beta\cosh{t}}dt = K_0(\beta)$ yields
\begin{equation}
    I_E(a; y_0) = \frac{\varepsilon P_0\sigma (E)}{\pi}\int_{-a}^{a} K_0\left( \frac{|y-y_0|}{\lambda}\right) dy
\end{equation}
where $K_0$ is the modified Bessel function of the second kind~\cite{Gradshteyn2014}. 

Next assuming $y_0 > a$ (i.e., the PhC is outside the slit), solving the integral in $y$ yields the following final expression
\begin{align}
    I_E(a; y_0) &= \frac{\varepsilon P_0 \sigma (E)}{\pi}\lambda ( r_+\left[ K_0(r_+)L_{-1}(r_+) + K_1(r_+)L_0(r_+) \right] \\ &\quad - r_-\left[ K_0(r_-)L_{-1}(r_-) + K_1(r_-)L_0(r_-) \right]),
\end{align}
where $r_\pm = (y_0\pm a)/\lambda$, $K_{0,1}$ are the modified Bessel functions of the second kind, and $L_{0,-1}$ are the modified Struve functions~\cite{Gradshteyn2014}.

We use this model to fit the observed intensity data in experiment using just two fit parameters: $\lambda$ and $C(E) \equiv \varepsilon P_0 \sigma(E)/\pi$. The results are shown in Fig.~4 of the main text. The good quantitative agreement between experiment and our model indicates that remote collection phenomenon is well-described by a minimal two-parameter framework consisting of an effective photon attenuation length and an energy-dependent collection scale. Because the model assumes only uniform isotropic emission, cylindrical spreading, and exponential loss, the fit is consistent with a simple geometric picture for the propagation of photons, rather than resonant or near-field nanophotonic effects, while nanophotonic patterning primarily modifies the local volumetric out-coupling efficiency.

\bibliographystyle{ieeetr}
\bibliography{references}